 \documentclass[acmsmall,screen]{acmart}

\usepackage{booktabs} 
\usepackage{subfig}
\usepackage{hhline}
\usepackage{array}
\usepackage{afterpage}
\usepackage{placeins}
\usepackage{color}
\usepackage[nolist,nohyperlinks,smaller]{acronym}
\usepackage{caption}
\usepackage{tabularx}
\usepackage{threeparttable}

\begin{acronym}
    \acro{USD}{US Dollars}
\end{acronym}

\def\changed#1{{\color{black}#1}}

\newcommand{\lei}[1]{\textcolor{black}{#1}}
\newcommand{\tc}[1]{\textcolor{black}{#1}}
\newcommand{\os}[1]{\textcolor{black}{#1}}
\newcommand{\sys}{Auggie}

\newcommand{\bear}{auggie}

\AtBeginDocument{%
  \providecommand\BibTeX{{%
    \normalfont B\kern-0.5em{\scshape i\kern-0.25em b}\kern-0.8em\TeX}}}

\setcopyright{acmcopyright}
\copyrightyear{2022}
\acmYear{2022}

\begin{document}

\title{Auggie: Encouraging Effortful Communication through Handcrafted Digital Experiences}


\author{Lei Zhang}
\authornote{All authors contributed equally to this research.}
\email{raynez@umich.edu}
\affiliation{
    \institution{University of Michigan}
    \city{Ann Arbor}
    \state{MI}
    \country{USA}
}

\author{Tianying Chen}
\authornotemark[1]
\email{tianyinc@andrew.cmu.edu}
\affiliation{
    \institution{Carnegie Mellon University}
    \city{Pittsburgh}
    \state{PA}
    \country{USA}
}

\author{Olivia Seow}
\authornotemark[1]
\email{oliviaseow@gmail.com}
\affiliation{
    \institution{Massachusetts Institute of Technology}
    \city{Cambridge}
    \state{MA}
    \country{USA}
}

\author{Tim Chong}
\email{tchong@snapchat.com}
\affiliation{
    \institution{University of Washinton}
    \city{Seattle}
    \state{WA}
    \country{USA}
}

\author{Sven Kratz}
\email{skratz@snapchat.com}
\affiliation{
    \institution{Snap Inc.}
    \city{Seattle}
    \state{WA}
    \country{USA}
}

\author{Yu Jiang Tham}
\email{yujiang@snap.com}
\affiliation{
    \institution{Snap Inc.}
    \city{Seattle}
    \state{WA}
    \country{USA}
}

\author{Andrés Monroy-Hernández}
\email{amh@snap.com}
\affiliation{
    \institution{Snap Inc.}
    \city{Princeton}
    \state{NJ}
    \country{USA}
}

\author{Rajan Vaish}
\email{rvaish@snap.com}
\affiliation{
    \institution{Snap Inc.}
    \city{Santa Monica}
    \state{CA}
    \country{USA}
}

\author{Fannie Liu}
\email{fannie@snap.com}
\affiliation{
    \institution{Snap Inc.}
    \city{New York}
    \state{NY}
    \country{USA}
}







\begin{abstract}
Digital communication is often brisk and automated. 
From auto-completed messages to ``likes,'' research has shown that such quick and easy interactions can affect perceptions of authenticity and closeness. On the other hand, effort in relationships can forge emotional bonds by conveying a sense of caring, and is essential in building and maintaining relationships. To explore \textit{effortful} communication, we designed and evaluated \sys{}, an iOS app that encourages partners to create digitally handcrafted AR experiences for each other. 
Auggie is centered around crafting a 3D character with photos, animated movements, drawings, and audio for someone else. 
We conducted a two-week-long field study with 30 participants (15 pairs), who used \sys{} with their partners remotely.
Our qualitative findings show that \sys{} participants engaged in meaningful effort through the handcrafting process, and felt closer to their partners, although the tool may not be appropriate in all situations.
We discuss design implications and future directions for systems that encourage effortful communication.
\end{abstract}

\ccsdesc[500]{Human-centered computing~Empirical studies in HCI}
\ccsdesc[500]{Human-centered computing~Empirical studies in collaborative and social computing}
\keywords{effort, communication, augmented reality}

\begin{teaserfigure}
  \includegraphics[width=\textwidth]{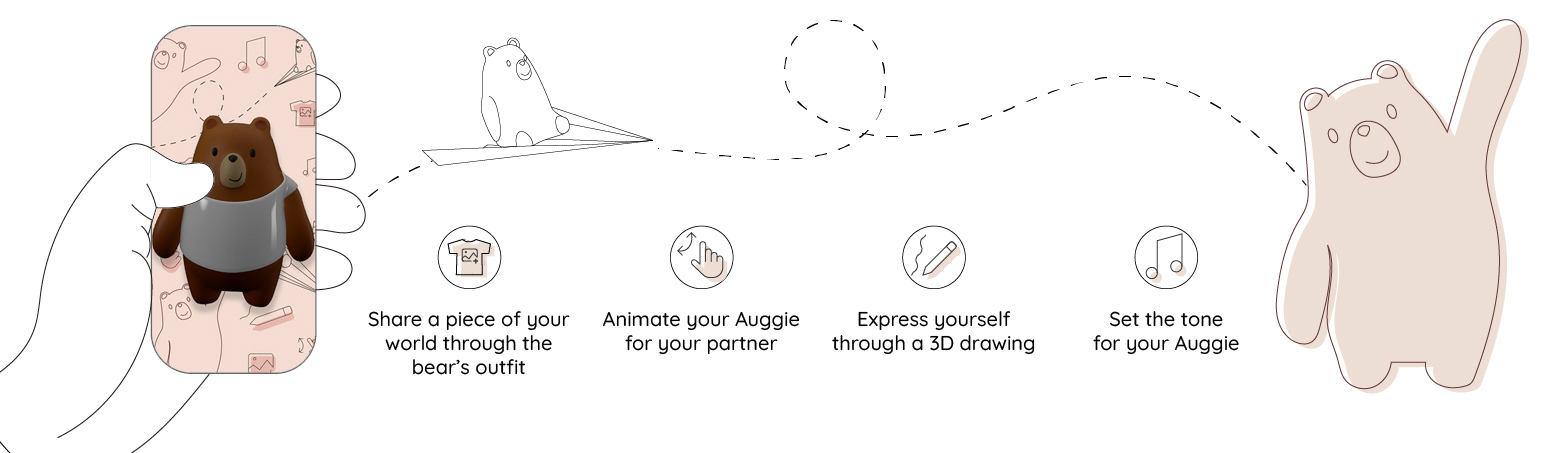}
  \caption{Description of Auggie, an iOS app that enables partners to create handcrafted AR experiences for each other, based on customizing a 3D character.}
  \label{fig:01_teaser}
\end{teaserfigure}

\maketitle

\section{Introduction}
The strategies originally developed to help computer users rapidly navigate the digital world~\cite{bao2006fewerclicks} have strongly influenced the design of digital social platforms. For instance, single-click reaction buttons frequently accompany or replace prompts for free response that could include personalized voice, text, and images~\cite{scissors2016s}. Autocomplete features are also increasingly used, wherein fixed presets are selected over custom expressions~\cite{doube2012autocomplete}. While low barrier-to-action digital interaction techniques can allow users to quickly initiate social interaction, too little effort can lead to concerns around authenticity~\cite{monroy2011computers}, trust~\cite{jakesch2019ai}, and feelings of support~\cite{wohn2016affective, spottswood2019beyond}. These are critical elements of \textit{meaningful} social connections, which involve emotional impact that can improve the quality of people's lives and relationships~\cite{litt2020meaningful}. For example, ``like’’ buttons are less associated with well-being improvements than more \textit{effortful} counterparts, including tailored expressions of support~\cite{hallinan2018like, burke2016facebook}. Moreover, both researchers~\cite{lessonsinloneliness} and popular media~\cite{vega_2017, fleishman_2018} have noted that simply triggering more interactions does not necessarily lead to more meaningful interactions.

To encourage meaningful digital interactions, we draw inspiration from the analog world. People express genuine affection by putting in effort to handcraft letters, scrapbooks, and gifts~\cite{lindley2009desiring, fuchs2015handmade}. By virtue of the medium alone, the effort in analog expressions translates relatively well to the receiver. Handwritten letters, for example, are perceived as more effortful than typing a long email, even if they take the same amount of time and mental energy: they are unique, personal, physically written and received, and, if properly cared for, even ``last centuries''~\cite{riche2010hard}. Better yet, if the creation process is documented and conveyed along with the analog gift, they can evoke even more meaningfulness~\cite{rosner2010spyn}. 

Yet, as digital communication becomes increasingly mainstream, handwritten letters and handcrafted scrapbooks might feel tedious and antiquated, while modern lightweight and automated communications can feel impersonal in comparison. Research investigating effort-provoking forms of digital communication, such as lengthy text messages, has had limited results, and findings show that effort ``can be resented just as easily as it could be valued, depending on the manner in which it arises''~\cite{kelly2018s}.

We attempt to address this by lowering procedural barriers to make effort more approachable, designing a communication system that enables versatile, personal expressions of effort. In this research, we study the feasibility of encouraging effortful digital communication and its ability to facilitate meaningful remote interactions. We designed and evaluated \sys{}, an iOS app that encourages partners to create digitally handcrafted Augmented Reality (AR) experiences for each other (see Fig.~\ref{fig:01_teaser}). \sys{} encourages effortful communication by making handcrafting for someone playful, personal, and engaging. Each experience, or \textit{\bear{}}, contains a 3D character that people can send to partners, along with custom photos, animated movements, drawings, and audio. We designed Auggie's interface and easy-to-use features to reduce the tedium and perceived expertise associated with handcrafting while enabling a wide range of creative possibilities for custom expressions and storytelling. All \bear{}s also contain a ``behind-the-scenes'' view where recipients can see the process a sender took to craft it for them, and thus, the effort they put in. We chose mobile AR because it offers immersive qualities~\cite{scholz2016immersive} on ubiquitous personal devices, giving us a widely accessible means to approximate and build on the experience of analog handcrafting on a digital medium.

This paper describes the design and system architecture of Auggie, along with results from a two-week-long field deployment with 30 participants (15 pairs) who used \sys{} remotely with close others. Our results show that, on aggregate, \sys{} successfully encouraged participants to put effort into crafting visual narratives beyond what is possible in physical reality. Participants found themselves inspired to digitally handcraft supportive, funny, and authentic expressions, which helped facilitate meaningful connection and feelings of presence. At the same time, this research also uncovered some unresolved challenges of effort, such as creativity blocks and appropriateness in certain situations.

Taken together, this paper contributes: 1) the design and implementation of an effortful communication system using digital handcrafting in AR; 2) results from a two-week field study that evaluated this system's potential to encourage effort on a digital medium through handcrafting and create meaningful interactions; and 3) implications for future design and research on fostering effortful communication in relationships.

\section{Background}

\subsection{Defining Effort}
Effort is sometimes referred to as ``the work...done to cope with and overcome demand on physical and mental capacities''~\cite{kelly2018s}. However, in interpersonal contexts, defining effort in only this mechanical sense does not capture the nuances of human connections, including the meaningfulness of effort for the people involved~\cite{kelly2018s}. For instance, researchers have identified the importance of effort in relationship maintenance, i.e., the strategies that people employ to maintain and enhance their relationships, such as cheerful attempts to make interactions enjoyable and mutual disclosure~\cite{canary2015relationship, canary1992relational, stafford1991maintenance}. Markopoulos breaks down the effort invested in communication within these relationships, particularly over technology, into \textit{procedural} effort and \textit{personal} effort. 
\textit{Procedural} effort is the effort required to complete a task (e.g., opening a messaging app), which aligns with the traditional mechanical perspective. On the other hand, \textit{personal} effort is what makes the task special for the recipient (e.g., thinking about something to say that they would like). Personal effort is usually more appreciated than procedural effort~\cite{markopoulos2009design}, but procedural effort can also be valuable when it reflects the personal effort involved (i.e., one's willingness to take on procedural tasks)~\cite{kelly2017demanding}.

 Kelly et al. further categorized different qualities of personal effort that people value in social interactions: (1) going out of one's way for the recipient of their effort, (2) crafting for them, (3) overcoming personal challenges for them, (4) devoting time for them, and (5) accounting for them and their beliefs in communication with them~\cite{kelly2017demanding}. While people can engage in these qualities within existing communication channels (e.g., sending a personalized text message), few prior researchers and social platform designers have \textit{intentionally} designed for these qualities. We still have limited understanding on both how to encourage meaningful, personal effort in digital communication systems, and its subsequent effects on relationships. The present research aims to address this and expand on the design space for effortful communication, exploring the potential to integrate meaningful qualities of effort, specifically ``crafting,'' ``devoting time,'' and ``accounting for the recipient and their beliefs,'' into communication technology.
 
\subsection{Supporting Personal Effort in the Digital Realm}
Though HCI researchers have acknowledged the need for communication systems to incorporate effort to improve and enhance relationships~\cite{king2007slow,lindley2009desiring,riche2010hard}, few works have explored this space. In one recent work, researchers created a physically effortful device to browse one's Twitter feed -- a hand crank that one must turn to scroll. They found that with the manual scroll, people focused more on the meaningfulness of the posts they read, while also becoming more aware of their Twitter usage~\cite{song2021crank}. Although researchers built this project using a social platform, the user's effort led to an individual rather than an interpersonal outcome. Moreover, this research increased the \textit{procedural} effort, rather than the \textit{personal} effort, of a task.

Effort explorations in direct interpersonal communication have primarily focused on text messaging. ``Message Builder,'' for example, is a text-based messaging app that encourages exchanging increasingly long messages. 
However, the authors found that while effortfully written lengthy messages produced some meaningful interactions, participants were frustrated by the inconvenience of the required procedural effort to access the system and read/write messages~\cite{kelly2018s}. This work highlights the need to balance design goals with users' expectations when increasing procedural effort in a task like text-based messaging. Other research examples in this space include ``Lily''~\cite{loveinlyrics2019}, which encourages people to manually revise text messages based on lyrical inspirations, and ``DearBoard''~\cite{dearboard2021}, where communication partners can collaboratively customize message keyboards, though increasing meaningful effort was not the primary goal of either system.

Our research expands on text-based messaging to explore new modes and opportunities for designing around \textit{personal} effort in communication. In particular, we focus on taking an analog task known to require a high level of \textit{personal} effort, and make it less \textit{procedurally} effortful. We do this through \sys{}, a playful system focused on the crafting quality of effort~\cite{kelly2017demanding}, engaging people in effort through digitally handcrafting gifts. As opposed to ``digital craftsmanship'', where a craftsperson may use software to aid in the fabrication of a physical item~\cite{tobiasdigitalcraftsmanship}, our use of ``digital handcrafting'' refers to the process of investing personal effort to produce a \os{digital gift-like experience for another person. In the present work, we use AR as the medium for digital handcrafting.}

We chose to focus on the effort conveyed by ``crafting for others'' as handicrafts are often perceived as authentic and ``made with love,'' which contributes to their value and attractiveness~\cite{frizzo2020genuine,fuchs2015handmade}. Love and authenticity are key elements of interpersonal relationships~\cite{brunell2010dispositional}. Therefore, it is not surprising that handcrafting something for another person contributes to relationship maintenance~\cite{minowa1999love}. Moreover, gifting handicrafts has long been a key way to convey personal effort~\cite{stafford1991maintenance}. By investing effort into creating something personal, the craft giver is often viewed as actively investing effort into the relationship~\cite{minowa1999love}. More uniquely, a handcrafted artifact highlights two layers of meaningfulness: in the end product itself, and the process that the creator took to bring it to life~\cite{luutonen2008handmade}. For example, when knitting a sweater, the creator may enjoy the process of picking out the yarn, measuring the size, and finishing and wrapping up the gift as a whole experience that represents their effort towards the relationship~\cite{minowa1999love}. The final handcrafted product is imbued with love from the creator's hands, which personally and physically interacted with it~\cite{fuchs2015handmade}.

In the present work, we investigate whether the meaningfulness of physically handcrafting for others can effectively translate to the \textit{digital} realm via an AR representation. We study how we can encourage personal effort through handcrafting in a digital communication tool. We explore digital handcrafting as a means to facilitate meaningful social interactions, and employ a qualitative approach to understand its potential for encouraging effort through \sys{}, a mobile application for handcrafting AR experiences. By evaluating \sys{}, we derive design implications for future researchers and practitioners to create digital systems that provoke personal effort to improve interpersonal outcomes.

\section{Preliminary Study}

\tc{To guide our design of an effortful digital communication system, we first sought to understand how effort is currently conveyed online. 
Prior research has provided some evidence on how people invest effort in their online interactions (e.g., carefully written text messages or social media posts)~\cite{kelly2017demanding}. We expand on this work by surveying people more broadly to understand their perceptions of effort on a wide range of digital platforms. We aimed to narrow in on characteristics of online effort that could inform our design direction, such as the frequency at which people engage in it and its current limitations.
We use this preliminary study as a first step to define a set of design goals for our system.}

We conducted a survey with 70 participants \tc{who spoke fluent English} using the online participant pool Prolific\footnote{https://prolific.co/}. \tc{Participants were compensated through Prolific for completing the survey following Prolific's recommended compensation guideline\footnote{https://researcher-help.prolific.co/hc/en-gb/articles/4407695146002-Prolific-s-payment-principles}.}
Participants were asked to describe the last time they put in extra effort for someone else online (as a ``sender'' of effort), the last time someone else put in extra effort for them (as a ``receiver'' of effort), \changed{and if there was anything they wished was different about those situations. They were also asked about their relationship with the people they mentioned, and how often they put in extra effort for them and vice versa.} After removing extremely short responses (e.g., one word answers for every question), we obtained 49 valid responses. \tc{The remaining participants included 35 male, 13 female, and 1 non-binary person, with ages ranging between 18 to 50 (M=25.02 and SD=7.29). 74\% of participants identified as White, 10\% as Hispanic, Latino, or Spanish, 8\% as Black or African American, 4\% as Asian, and 4\% as mixed or multi-racial (self-described as White/Asian and Caucasian/African-American).}

We analyzed responses to understand when, with whom, and how people engage in effortful communication online and its current limitations. 
We open coded the responses according to the effort qualities proposed by Kelly et al. ~\cite{kelly2017demanding}, as well as the context (e.g., planned or unplanned) and social goals (e.g., comfort, celebration, thinking of you). Two coders coded a subset of the data independently until they reached 80\% agreement, and then independently coded the rest of the data. \tc{Participants most commonly reported effort conveyed between close friends (38.8\%), followed by spouse and significant others (20.4\%), friends (20.4\%), family members (10.2\%), strangers online (8.0\%), and acquaintances (2.0\%)}.
The following insights from the survey guided the design of \sys{}:

\paragraph{Personalization is the most memorable type of effort}
The most memorable effortful interactions for both senders and receivers are those where the sender tailored the end product to the receiver's interest. \changed{This aligns with the ``accounting for the recipient and their beliefs'' quality of meaningful effort described in prior work~\cite{kelly2017demanding}.}
The most common forms for this were gifting, sharing memories, and music.
For gifting, senders bought gifts online (e.g., an online game, model kits) that they knew the receiver would enjoy.
Shared memories were strengthened by exchanging photos and videos of such experiences. 
Senders would also put effort into personalizing music for the receiver. 
For instance, one respondent said their friend sang their favorite song \changed{over a call} to them to cheer them up. Based on this finding, we conceptualize a system that allows for personal effort through high levels of customization~\cite{markopoulos2009design}.

\paragraph{Effort is demonstrated day-to-day, not just on special, planned occasions}
We found that most effortful digital interactions occurred in everyday occasions (e.g., sending sentimental songs when one party was \textit{thinking of the other}), while only 30\% and 15\% of responses for effort conveyed and received, respectively, were for special, planned occasions (e.g., sending a recorded video message on a birthday). 
Based on this result, we believe that effortful communication should not only be designed for special occasions, but also be applicable on day-to-day occasions. As such, we designed the system to support versatile crafting, where people have the choice to craft simple (e.g., just saying hi) or complex (e.g., elaborate stories) content that they can contextualize for different occasions with images, audio, and animations.

\paragraph{Sense of presence is lacking in remote effort} 
Participants highlighted a deep desire to feel a greater sense of presence from their remote partners. 
This aligns with prior literature on social presence, or ``the sense of being with another,'' which is well-known to be missing in remote communication~\cite{biocca2003toward}.  Respondents remarked that they particularly appreciate the effort from remote partners in attempting to be ``present'' through video chats, voice calls, or simply being available on a digital platform \changed{(e.g., a participant noting a friend who was ``with them'' on a Discord call for several hours to cheer them up).} However, there is still a clear demand for an increased sense of tangible presence in any form of remote communication. \tc{For example, the aforementioned participant who described their friend singing to them over a call continued to say, ``I would like her to come to my place so I can hug her but she couldn't cause we live in different cities.''} To address this, we incorporate AR in our system to make use of its immersive qualities and ability to simulate physical presence~\cite{scholz2016immersive}, further described in Section \ref{presence}.

\os{Based on these findings, our research team underwent several rounds of brainstorming to generate ideas for an effortful communication system. This involved each team member individually presenting ideas, followed by group discussion and iteration, and finally narrowing down on specific concepts that we felt best aligned with the design insights from our preliminary study and prior work. We also built and user tested several low-fidelity prototypes to further develop our ideas. We describe the final system design in the following section.}



\section{Auggie System}
\sys{} is an iOS app that allows partners to send digitally handcrafted AR experiences (i.e., \bear{}s) to each other in the form of an animated 3D bear. The bear character was inspired by teddy bears, which are common gifts among close partners. The character also facilitates the creation of expressive and embodied animation for shared narratives, which can increase feelings of personalization and presence. This represents one implementation by which we can study the impact of effort conveyed through an activity typically of high \textit{procedural} effort (in this case handcrafting, which usually require time, material costs, and skill) made more \textit{personal} and approachable.

Senders can craft an AR experience by animating the 3D bear, as well as personalizing its outfit and adding 3D drawings, background music, and voice notes. The bear then ``travels'' via an AR paper airplane to its designated recipient, who can view the AR experience in their own physical environment. The recipient can also view the sender's ``behind-the-scenes'' creation process, which is automatically compiled by the system.

\subsection{Situating Auggie in the Physical World}\label{presence}
\lei{\sys{} integrates AR by enabling users to point their phone camera to a surface in their physical environment to craft or view the 3D bear and other elements of the experience.
We chose AR as a medium for crafting for several reasons.
First, creating and interacting with virtual objects in AR could stimulate the sense of handcraftedness, since a user is personally and physically interacting~\cite{fuchs2015handmade} with an object that feels real~\cite{kroupi2014predicting, higuera2017psychological}.
We also aim to make the handcrafted digital experience more engaging using AR due to its unique and promising potential to facilitate playfulness~\cite{dagan2022project} and users' creativity~\cite{zund2015augmented}. 
In addition, the physical and spatial affordances of AR could enable a heightened sense of presence~\cite{scholz2016immersive}, which is often lacking in remote effortful communication, as found in our preliminary study. 
By overlaying the virtual bear on users' physical environments, Auggie could provide a subjective sense of a partner ``being present'' in the experience, where the bear represents the partner in one's space.
}
Finally, AR can enhance expressiveness by enabling users to imagine and create experiences for each other that are otherwise not possible in the physical world. 
The ability of AR to superimpose virtual information onto the physical world has enabled experiences beyond the physical world such as storytelling~\cite{bai2015exploring}, virtual structure building~\cite{guo2019blocks}, and museum touring~\cite{damala2008bridging}.
\lei{We aim to leverage the expressiveness and versatility of AR experiences to enable high levels of personalization in Auggie, which our preliminary study suggests is the most memorable type of effort. 
Altogether, we introduce a unique avenue for handcrafted digital experiences.
}

\subsection{Crafting Personalized Auggies}

\begin{figure}
     \centering
     \subfloat[Outfit\label{fig:component_outfit}]{\includegraphics[height=0.4\textwidth]{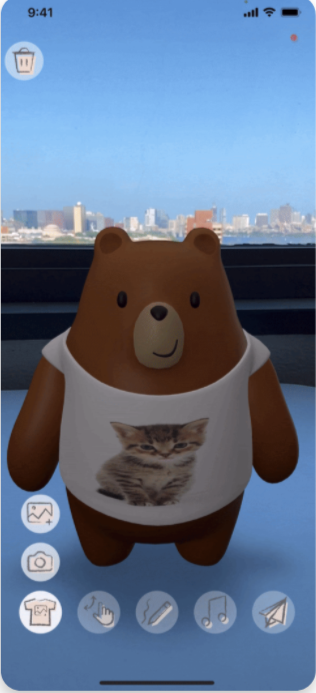}} 
     \hfill{}
     \subfloat[Animation\label{fig:component_oanim}]{\includegraphics[height=0.4\textwidth]{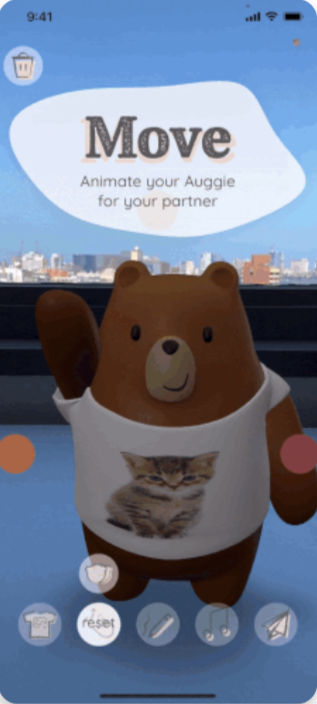}}
     \hfill{}
     \subfloat[3D Drawing\label{fig:component_draw}]{\includegraphics[height=0.4\textwidth]{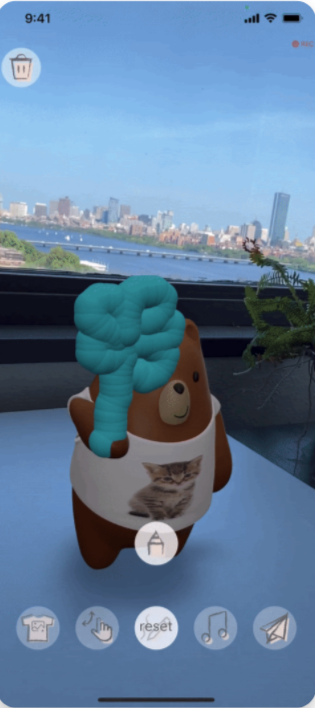}} 
     \hfill{}
     \subfloat[Background Music\label{fig:component_bgm}]{\includegraphics[height=0.4\textwidth]{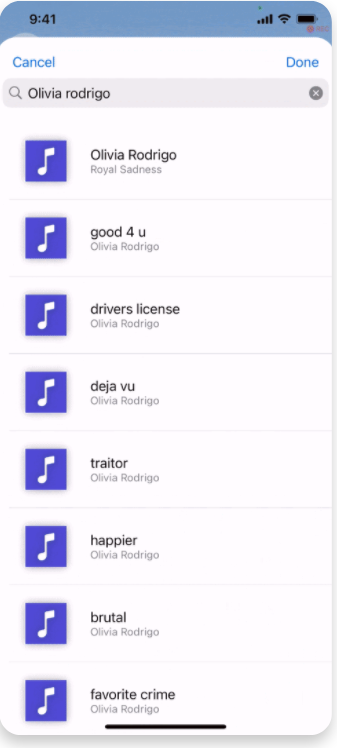}} 
     \hfill{}
     \subfloat[Voice note \& Confirm\label{fig:component_voice}]{\includegraphics[height=0.4\textwidth]{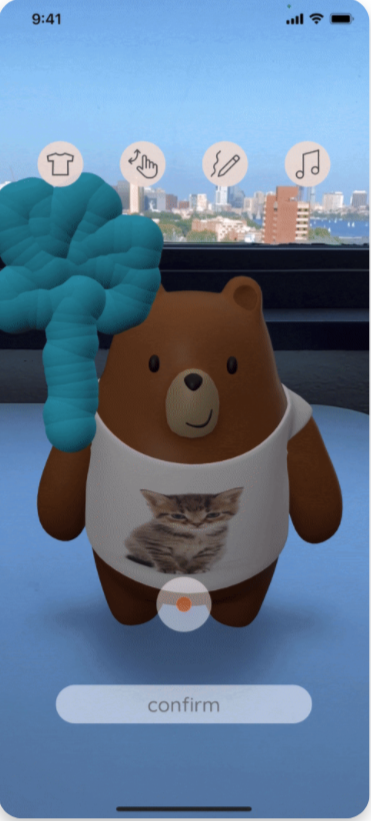}} 
     
     \caption{Components of crafting personalized \bear{}s. Users can personalize each component of an \bear{} by selecting from a set of buttons on the bottom - (a) when \textit{outfit} is selected, users can attach a picture to the bear's t-shirt by selecting from the image gallery or taking a photo; (b) when \textit{animation} is selected, users can author the animation of the bear or toggle the partner bear button to animate two bears together; (c) users can create a 3D drawing by holding the pen button and moving the phone in the physical world; (d) when \textit{music} is selected, users can search and select a background music for the \bear{}; (e) while reviewing the experience and before sending it, users can add a voice note by pressing the record button.}
     \label{fig:system}
\end{figure}

Our survey results and prior work both indicate that people value effort in personalization, such as of gifts, photos, and music.
\sys{} encourages users to engage in such effortful communication through crafting personalized AR experiences (or \bear{s}) for another person, as if creating a gift. To support different occasions and desired levels of effort, users can personalize the experience as much or as little as they would like, where they could even send ``blank'' \bear{}s that simply show the bear.
The personalization process consists of five components: outfit, animation, 3D drawing, background music, and voice note (as seen in Fig. \ref{fig:system}).

\subsubsection{\changed{Outfit, music, and voice note}}
Our preliminary study revealed that people appreciated the effort made to curate photos and hand select music for them.
On \sys{}, users can select from the photo gallery or take a photo (Fig. \ref{fig:component_outfit}) to personalize the bear's outfit. 
Selecting from existing photos allows users to engage in digital handcrafting with shared memories or shared interests (e.g., inside jokes).
Taking a photo can also draw a connection between the digital and physical worlds, allowing users to integrate elements from the physical into the digital world and potentially feel more present with each other~\cite{memento2007}, further addressing concerns described in the preliminary study. 
Users can also set the mood of the experience by selecting background music from the iTunes music library (Fig. \ref{fig:component_bgm}).
In addition to adding existing songs, users can attach a voice note to incorporate more personal and expressive audio (Fig. \ref{fig:component_voice}). We included these audio options based on our preliminary study results and prior work, which shows that music influences positive emotion, engagement, and relationships~\cite{croom2015music}.

\subsubsection{Animation} We included animation to enable a wide range of expressions (e.g., waving, sad, jumping) while fostering a sense of handcraftedness. 
Animations and animated images are frequently used in the digital communication for storytelling and self-expression through the forms of GIFs, emojis, and memes~\cite{griggio2019customizations}. 
These formats not only facilitate shared expressions and address the lack of facial and vocal exchanges in text-based communication channels, but also serve as important strategies to support connectedness and relationship building~\cite{sugiyama2015kawaii,hsieh2017playfulness}. Users animate the 3D bear's joints by dragging three circles on the screen corresponding to the bear's spine, left arm, and right arm (Fig. \ref{fig:component_oanim}).
Users can also make the bear jump using a single finger vertical swipe, and rotate the bear using a single finger horizontal swipe. The bear can also walk in the physical environment, moving when a user taps on a destination on the phone screen. To further increase perceptions of presence, we also included a feature where users can toggle the appearance of a partner or companion bear, whose actions are synchronized with the main bear when they are in close proximity, enabling effects like dancing or hugging.

In other digital animation tools like Autodesk Maya and Cinema 4D, animation is an investment of time and procedural effort in many aspects, such as learning how to model 3D objects, encode 3D movements, and set up lighting effects. Comparatively, our system has a small set of intuitive controls, aiming to reduce the amount of procedural effort required to produce aesthetically pleasing digital animations.
At the same time, users are able to create animations that are more complex and personal compared to selecting from a fixed library of generic emojis or animated stickers.
By creating animations from scratch in their physical environment, users can invest and communicate effort that feels handcrafted and personal.


\subsubsection{3D drawing}
Drawing is known to be an expressive way of creating personal and unique experiences in AR, such as for performances~\cite{saquib2019interactive} and storytelling~\cite{leiva2020pronto}.
In \sys{}, users can press the draw button and move the phone freely in their physical space to create 3D drawings (Fig. \ref{fig:component_draw}).
We incorporate this feature to allow further personalization as well as the ability to interact with the physical space.
This feature also enables users to create virtual props (e.g., a hat for the bear) as part of the handcrafted AR experience.

\subsection{Sending and Receiving \sys{}s}

\begin{figure}
     \centering
     \subfloat[Send-off Scene\label{fig:sendoff_scene}]{\includegraphics[height=0.4\textwidth]{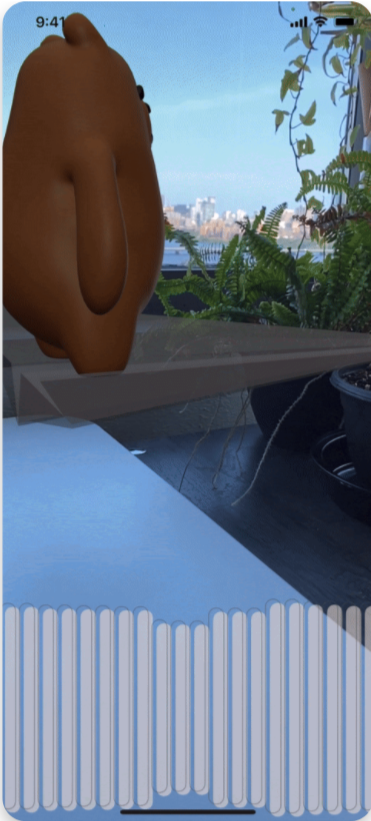}} 
     \hfill{}
     \subfloat[Viewing \sys{} in AR\label{fig:sendoff_viewing}]{\includegraphics[height=0.4\textwidth]{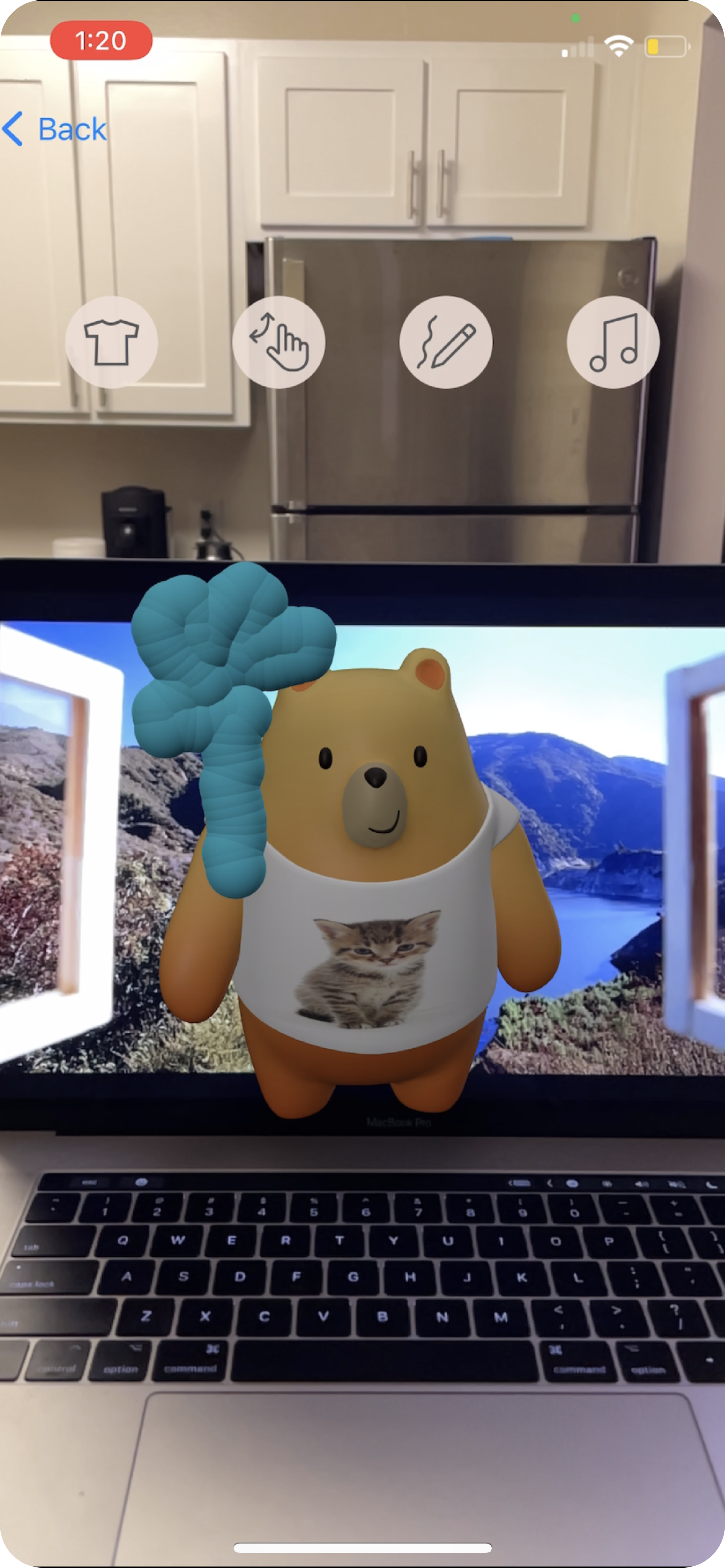}} 
     \hfill{}
     \subfloat[``Behind-the-scenes''\label{fig:sendoff_bts}]{\includegraphics[height=0.4\textwidth]{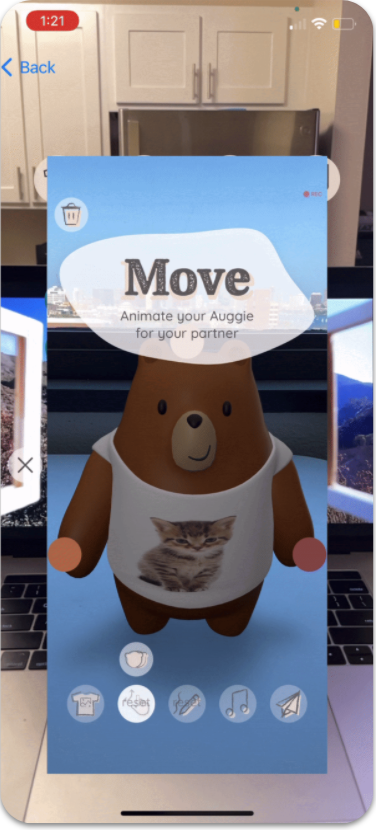}} 
     \hfill{}
     \subfloat[\lei{Inbox}\label{fig:sendoff_inbox}]{\includegraphics[height=0.4\textwidth]{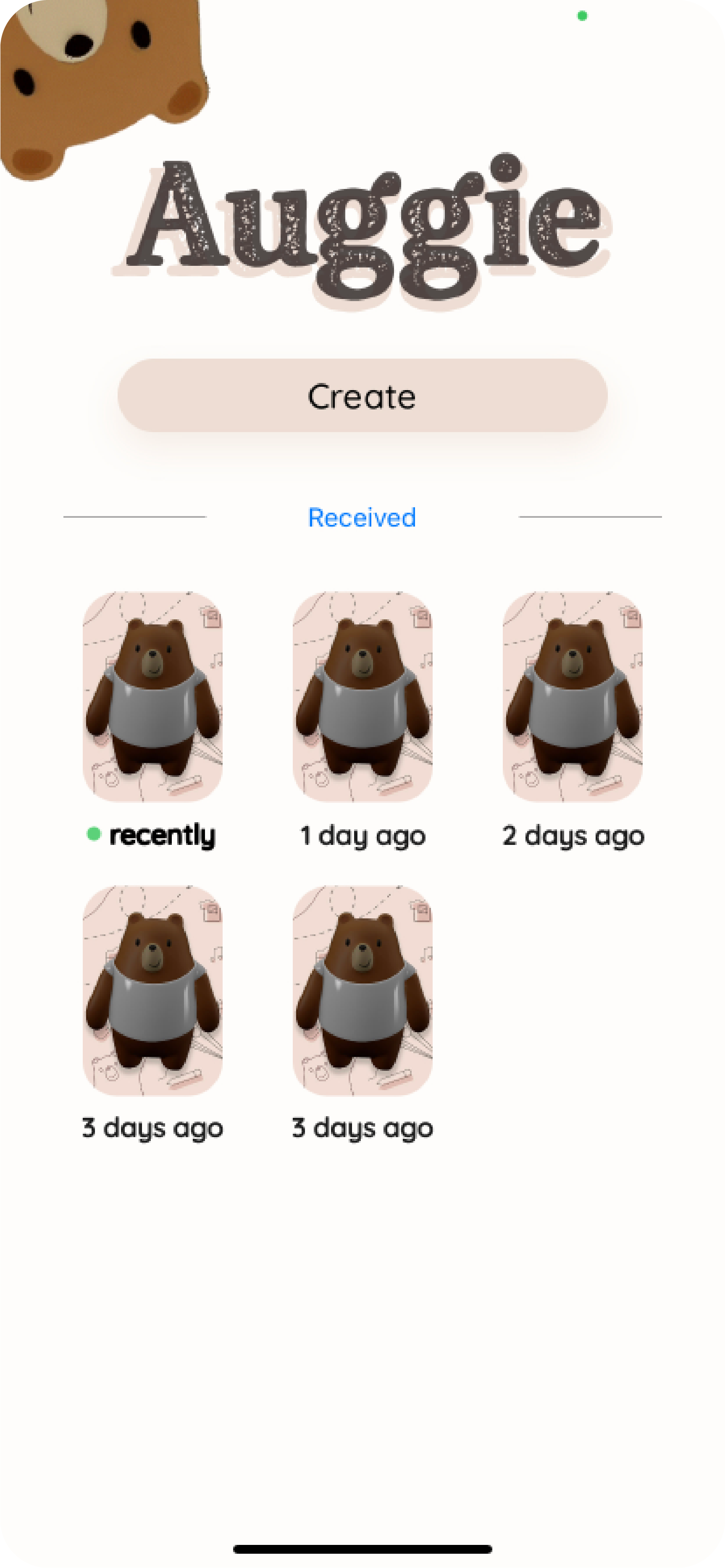}} 
     
     \caption{Sending and receiving \bear{}s: (a) after confirming, users can blow wind to the microphone to send off the bear and see the real-time audio wave; (b) receivers can view the \bear{} in AR and see available ``behind-the-scenes'' videos through the top buttons; (c) receivers can also click on one of the four categories (outfit, animation, drawing, and music) from the top buttons and see a screen recording pop-up of the sender's creation process for the selected category; \lei{(d) Received \bear{}s are saved in an inbox for users to rewatch.}}
     \label{fig:sendandreceive}
\end{figure}

\begin{figure}[h]
  \centering
   \includegraphics[width=0.5\columnwidth]{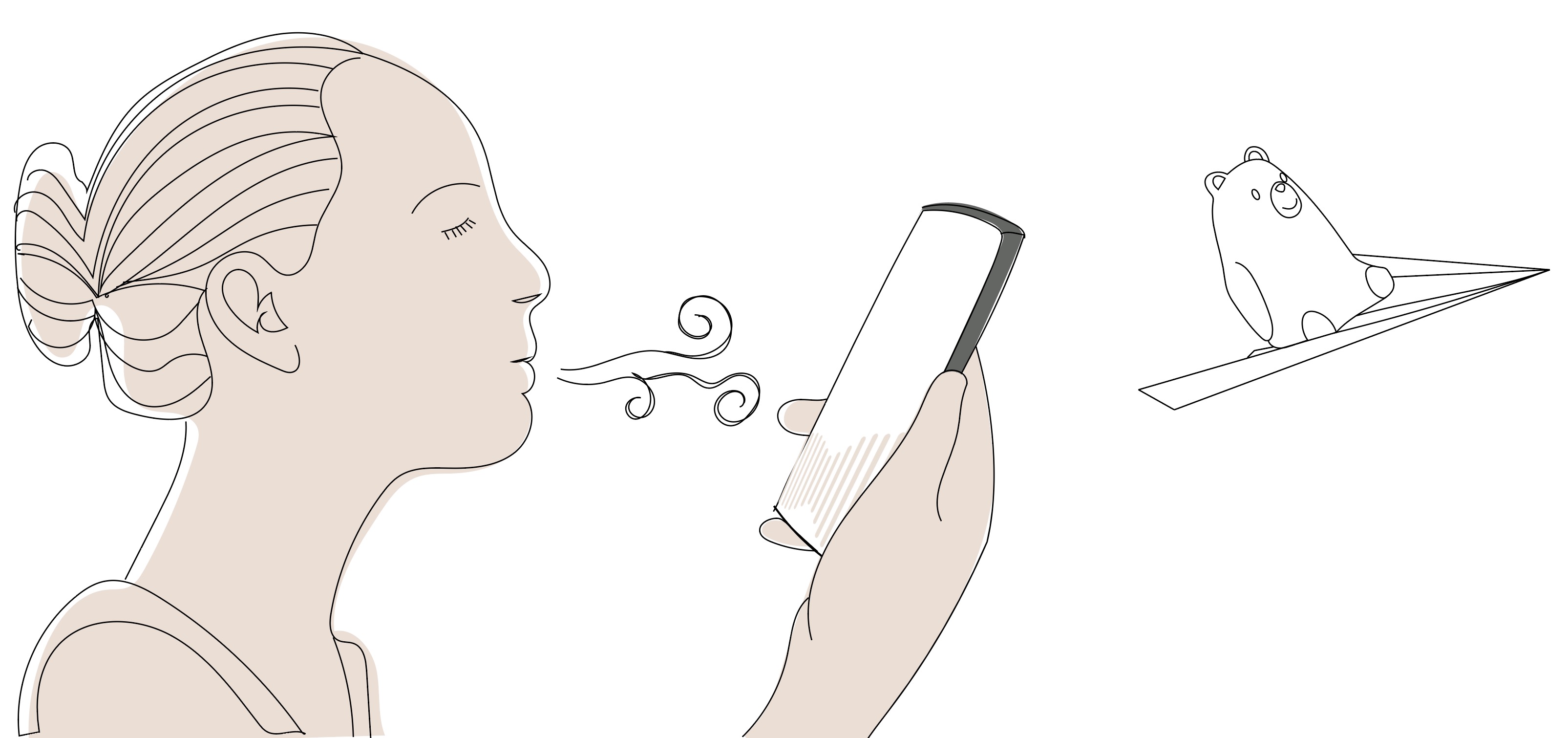}
   \caption{To send an \bear{}, users blow wind at their phone, which triggers an AR paper plane to fly into the distance, carrying a bear over to their partner.}
   \label{fig:blow_wind}
\end{figure}

After crafting a personalized \bear{} and before sending it, users may review the whole experience.
To send an \bear{}, users press the ``confirm'' button (as seen in Fig. \ref{fig:component_voice}) and view the bear climbing on a paper airplane in AR (see Fig. \ref{fig:sendoff_scene}). \changed{Users then need to blow wind into the phone's} microphone for the paper plane to take off (see Fig. \ref{fig:blow_wind}). 
When receiving an \bear{}, users can point the rear camera to a window or mid-air, and the paper airplane will fly into view. 
Users can then guide the airplane to land by guiding the camera to a surface. 
The \bear{} is then presented in the receiver's physical environment (see Fig. \ref{fig:sendoff_viewing}).
\lei{All the received auggies are saved on the phone and can be rewatched at any time through Auggie's inbox feature (see Fig. \ref{fig:sendoff_inbox}).}

Prior work highlights the importance of incorporating analog approaches to gifting. In particular, ``wrapping'' a gift in a physical object and mimicking the unwrapping ritual can harbor a sense of physical presence~\cite{kwon2017s}. Therefore, we intentionally incorporated interactions and visual representations similar to the physical world (i.e., blowing wind to make a paper plane fly, guiding its landing) into the sending and receiving process \os{(see Fig. \ref{fig:blow_wind})}.

\subsubsection{``Behind-the-scenes'' videos}
The perception of the effort one puts into a relationship is essential in determining the satisfaction that one has towards the relationship~\cite{stafford1991maintenance}. Additionally, demonstrating the creation process of a handmade gift and sending it alongside the gift itself can allow receivers to interpret the story behind the gift and increase meaningfulness~\cite{rosner2010spyn}.

However, the effort put into crafting a \textit{digital} artifact may not be easily deduced from the output. For example, a person may go through a lengthy process of trial-and-error to craft the artifact, which is lost in the final result. \lei{Prior work has thus explored several design concepts around revealing effort, centered primarily on text communication, and shown that information about effort can foster reflection and appreciation~\cite{kelly2017demanding, kelly2018designing}. To reveal effort in the realm of AR handcrafting,} \sys{} screen records the sender's ``behind-the-scenes'' creation process and automatically creates a timeline consisting of four videos (corresponding to outfit, music, animation, and drawing) that document the sender's effort (see Fig. \ref{fig:sendoff_bts}).
\lei{We chose to reveal effort through screen recordings since prior work has shown that replaying creation processes can demonstrate authenticity behind effort (i.e., whether the effort is invested willfully by the sender). That is, recordings can provide details behind effort such as pauses, which can help receivers ``think what senders are thinking'' and feel closer to them~\cite{kelly2018designing}.}
For privacy, we blur out aspects of the ``behind-the-scenes'' video that are external to the \sys{} app, such as photos in a user's gallery that may appear during the outfit selection process.



\subsection{Implementation}
\sys{} is implemented using SwiftUI\footnote{https://developer.apple.com/documentation/swiftui/}, 
RealityKit\footnote{https://developer.apple.com/documentation/realitykit/}, 
and ARKit\footnote{https://developer.apple.com/documentation/arkit}.
Messages are sent and received through HTTP requests and stored on an AWS\footnote{https://aws.amazon.com/} server.
Each message consists of six components: animation, 3D drawing, images, videos, music, and audio.
Animation is encoded using transforms of the \bear{} bear's skeleton on a frame-to-frame basis.
3D drawing is encoded using positions of spheres generated in each frame.
Images, videos, music, and audio are encoded in binary and uploaded to the server.
To replay a message on the receiver's end, the system downloads and decodes each component.
\section{Methods} \label{methods}
We conducted a field study with 30 participants (15 pairs) who used \sys{} for two weeks. We interviewed participants at the middle and end of the study in order to understand the details of their experience using \sys{}.

\subsection{Participants}

\setlength{\tabcolsep}{3.5pt}
\begin{table}
  \caption{Participant Demographics}
  \label{tab:participants}
  \small
  \begin{tabular}{lrlrlrlrlrlr}
    \toprule
    \multicolumn{2}{l}{\textbf{Pair IDs}} & \multicolumn{2}{l}{\textbf{Gender}} & \multicolumn{2}{l}{\textbf{Age}} & \multicolumn{2}{l}{\textbf{Occupations}} & \multicolumn{2}{l}{\textbf{Relationship}} & \multicolumn{2}{l}{\textbf{Location}} \\
    \midrule
    1a & 1b & M & M & 21 & 20 & Student & Student & \multicolumn{2}{l}{Friend} & Canada & Canada\\
    2a & 2b & F & F & 22 & 22 & Software Engineer & Student & \multicolumn{2}{l}{Friend} & US & US\\
    3a & 3b & F & M & 25 & 25 & Research Intern & Planning Manager & \multicolumn{2}{l}{Significant Other} & US & China\\
    4a & 4b & F & F & 33 & 30 & Marketing Manager & Account Director & \multicolumn{2}{l}{Friend} & US & US\\
    5a & 5b & M & F & 33 & 28 & Product Manager & Property Manager & \multicolumn{2}{l}{Sibling} & US & US\\
    6a & 6b & F & M & 26 & 28 & Student & Product Manager & \multicolumn{2}{l}{Significant Other} & US & US\\
    7a & 7b & F & F & 24 & 24 & UX Architect & Student & \multicolumn{2}{l}{Friend} & US & US\\
    8a & 8b & F & F & 25 & 25 & Student & Data Engineer & \multicolumn{2}{l}{Friend} & US & US\\
    9a & 9b & M & M & 30 & 26 & Student & Pharmacist & \multicolumn{2}{l}{Sibling} & US & US\\
    10a & 10b & F & F & 23 & 24 & Student & Bather & \multicolumn{2}{l}{Friend} & US & US\\
    11a & 11b & F & F & 22 & 22 & UX Lead & Technical Lead & \multicolumn{2}{l}{Friend} & US & US\\
    12a & 12b & F & M & 20 & 20 & Student & Student & \multicolumn{2}{l}{Significant Other} & US & US\\
    13a & 13b & F & F & 22 & 22 & Software Engineer & Equity Analyst & \multicolumn{2}{l}{Friend} & US & US\\
    14a & 14b & F & F & 26 & 20 & Student & Student & \multicolumn{2}{l}{Sibling} & US & India\\
    15a & 15b & F & F & 25 & 20 & Software Engineer & Student & \multicolumn{2}{l}{Sibling} & US & US\\
    \bottomrule
  \end{tabular}
\end{table}

We recruited participants from a technology company and four universities in the United States through Slack posts and e-mails.
Each participant was asked to select a partner with whom they are close but not currently cohabiting. \tc{We chose this population in order to investigate circumstances in which digital communication plays an essential role. By disallowing cohabiting participants, we focus on participants whose main method of communication is digital and reduce the impact that in-person interactions might have between the pairs.}
We recruited 50 participants (25 pairs); however, 10 pairs were removed from the study in the first week due to failing to meet the minimum study requirements (e.g., cohabitation status, not showing up to interview sessions, and technical issues such as incompatible iOS versions).
This left us 15 pairs of participants in total (n=30).
Table \ref{tab:participants} shows the demographic information of participants.
We compensated each participant with \$25 Amazon gift cards at the middle and end of the study, for a total of \$50 worth of gift cards.

\subsection{Procedure}
Participants first joined an hour-long onboarding session over Google Meet to receive study instructions. Prior to this session, they also completed a short entry survey about their background and relationship with their \sys{} partner. \tc{During the onboarding, participants were first introduced to the study and were informed of all the data that would be collected during the study. They were also informed that participation was voluntary and that they could drop out at any time without repercussion. All participants signed an informed consent.}
After obtaining consent, a member of the study team then introduced \sys{}'s functionalities and guided the participants through the process of creating an \bear{}. 
They ensured that participants understood how to use the application and were able to send and receive \bear{}s during the onboarding session. 

During the two-week study, the participants were asked to send at least one \bear{} to their partner daily. \tc{This instruction was not a strict requirement, but rather a means to encourage participants to send \bear{}s to each other. All participants were compensated fully regardless of the number of \bear{}s they sent, and were only excluded from our analyses in extreme cases (e.g., only sending one default \bear{} with no changes). While more freely using the app could potentially reveal how it fits naturally into participants' lifestyles, we encouraged minimal usage to ensure enough engagement during the study to inform our understanding of participants' experiences with the app.}

\changed{To understand participants' experiences, we asked participants to complete brief daily surveys and two semi-structured interviews during the study. The daily surveys} included open-ended questions about the \bear{}s that they sent and received, their perceptions of effort using \sys{}, and any technical issues they faced (see supplementary materials). We asked participants to fill out a minimum of three surveys per week.
After the first week, the participants completed a 30-60 min semi-structured mid-\os{study} interview over Google Meet, which included questions about participants' experiences creating and receiving \bear{}s, their perceptions of effort and AR while using \sys{}, and comparisons between \sys{} and other communication tools that they used (see supplementary material). The interviews were also used to clarify and follow-up on responses to their daily surveys for more detail. 
At the conclusion of the study, participants were offboarded and completed an exit interview and survey, which were similar to the mid-\os{study} interview (with additional questions about any changes in their usage and perceptions of \sys{}) and entry survey, respectively.
Throughout the study, we also recorded each user interaction in \sys{} with the corresponding user id, timestamp, and interaction event to track their overall usage of \sys{}. \changed{Our study proposal and protocol were reviewed internally by a panel consisting of independent privacy engineering and legal teams to ensure that the subject matter and approach complied with ethical standards and that participants’ data was processed appropriately.}

\subsection{Data Analysis}

We analyzed users' behavioral data and the transcriptions of the mid-\os{study} and exit interviews.
For the behavioral data, we parsed and cleaned the system event data and the message \tc{content} from the server, forming descriptive results of the usage of \sys{}.
For the qualitative analysis of the interview data, we took a grounded theory approach~\cite{strauss1998basics}. 
Two researchers first performed open-coding on three randomly selected participants' transcripts, identifying and labeling similarities across participants' experiences with \sys{} in order to develop a codebook.
Three researchers then refined the codebook on another subset of participant transcripts, individually coding the same transcripts.
The researchers validated the codebook on two additional participants' transcripts, meeting frequently to discuss disagreements. After achieving high inter-rater reliability, with a Krippendorff's $\alpha$ above 0.8, they divided and coded the rest of the transcripts in parallel.
Finally, we performed axial coding to group related codes into themes according to our research questions, which we refined during the paper writing process.
\section{Results}




\setlength{\tabcolsep}{3.5pt}
\begin{table}
  \caption{\changed{\sys{} Usage Data}}
  \label{tab:message}
  \small
  \begin{tabular}{lcccccc}
    \toprule
    \textbf{Pair IDs} & 
    \textbf{\# of Auggies} & 
    \textbf{\# of Animation} & 
    \textbf{\# of Voicenote} & 
    \textbf{\# of Drawing} & 
    \textbf{\# of Outfit*} & 
    \textbf{\# of Music}\\
    \midrule
    1a, 1b & 8, 8 & 8, 7 & 8, 5 & 7, 0 & 7, 5 & 5, 3\\
    2a, 2b & 8, 6 & 8, 6 & 4, 6 & 1, 0 & 0, 0 & 0, 0\\
    3a, 3b & 16, 17 & 11, 8 & 11, 17 & 5, 2 & 2, 11 & 0, 1\\
    4a, 4b & 12, 8 & 11, 8 & 9, 5 & 8, 3 & 3, 1 & 7, 5\\
    5a, 5b & 8, 11 & 6, 10 & 6, 4 & 1, 3 & 1, 4 & 0, 0\\
    6a, 6b & 6, 8 & 6, 8 & 6, 4 & 3, 2 & 4, 3 & 1, 1\\
    7a, 7b & 7, 9 & 7, 8 & 3, 9 & 0, 3 & 2, 1 & 1, 0\\
    8a, 8b & 14, 13 & 14, 11 & 5, 10 & 6, 5 & 2, 5 & 0, 2\\
    9a, 9b & 13, 7 & 9, 7 & 8, 1 & 4, 3 & 2, 0 & 0, 0\\
    10a, 10b & 17, 15 & 17, 12 & 12, 15 & 17, 14 & 16, 13 & 3, 2\\
    11a, 11b & 17, 3 & 15, 3 & 13, 3 & 4, 1 & 2, 1 & 6, 0\\
    12a, 12b & 11, 16 & 10, 16 & 1, 0 & 8, 16 & 4, 7 & 6, 4\\
    13a, 13b & 21, 25 & 20, 25 & 5, 9 & 5, 6 & 3, 0 & 5, 1\\
    14a, 14b & 8, 8 & 6, 4 & 0, 2 & 4, 5 & 2, 4 & 3, 3\\
    15a, 15b & 7, 10 & 7, 10 & 7, 4 & 2, 9 & 4, 9 & 1, 3\\
    \midrule
    \textbf{Mean} & 11.2 & 9.9 & 6.4 & 4.9 & 3.9 & 2.1\\
    \textbf{SD**} & 5.0 & 4.8 & 4.3 & 4.4 & 3.9 & 2.2\\
    \bottomrule
  \end{tabular}
  \begin{tablenotes}
   \item* Outfit: t-shirt image.
   \item** SD: standard deviation.
  \end{tablenotes}
\end{table}

Participants' experiences using \sys{} during the study suggest that they indeed engaged in effort while crafting animated AR experiences for their partner. 
\lei{As Table \ref{tab:message} shows, participants sent an average of 11.2 \bear{}s, with animations as the most frequently used component (10/11 \bear{}s on average), but the number of \bear{}s sent varied widely across participants. Based on the interviews, this was partly due to different perceptions of \sys{} usage scenarios, where some participants perceived everyday usage while others preferred it for special occassions. For instance, P13b, who sent 25 \bear{}s in total, described \sys{} as ``similar to Messenger or Snapchat where we're just sending little bits of our updates every day.'' On the other hand, P11b sent 3 \bear{}s in total, and described using it for special occasions ``like a birthday or an anniversary,'' rather than daily, viewing Auggie as ``mostly about creating something new'' and not ``sharing anything about [themselves].''}

Additionally, in line with our design motivations, we found that participants perceived how much effort their partners put into crafting through ``behind-the-scenes'' videos.
As one participant pointed out, with this feature, ``you know where this person is...you could kind of see if they put effort into making these \bear{}s or if they didn't'' (P6b).

In the following subsections, we provide more detailed insights into the participants' usage and perceptions of \sys{}. We center our findings around two major themes: 1) how digital handcrafting encouraged people to engage in effortful communication, and 2) the effects of handcrafting in the digital space.

\subsection{Encouraging Effort through Handcrafted Digital Experiences}
Overall, our results suggest that the design of \sys{} encouraged effort by promoting a sense of agency, creating the feeling of gift-giving, and enabling digital storytelling experiences through the process of crafting \bear{}s.
Participants felt they had control over the creation process, which encouraged them to invest effort into crafting unique personal experiences for their partner as if it was a gift. Participants also highlighted the ability to craft stories that allowed them to express themselves beyond the possibilities of the physical world.

\subsubsection{Agency over handcrafting encouraged effort}
Agency refers to the awareness and control over one's action~\cite{sep-action}. 
Participants reported having agency over crafting their \bear{}s. 
The design of \sys{} provided participants with a variety of crafting options such as animation to orchestrate a story, 3D drawing to embellish the environment, music to set the mood, and voice note to express themselves.
These elements together provided wide control over what the final experience would look like and further enabled users to craft experiences that feel personal:


\begin{quotation}
``It's a lot more personal compared to all the other apps that are out there, especially because you control the bear and because you're able to modify all of its functions and all the other features. It's basically, \textit{you} create it.''~-P2b
\end{quotation}

In addition, participants' agency over \bear{}s evoked a sense of accomplishment and ownership.
More specifically, participants felt proud of themselves after investing effort and using creativity in the crafting process, especially given that many of them did not have artistic backgrounds. Through this process, they could feel like \textit{creators}, and the experiences they created belonged to them:
\begin{quotation}
``I'm not an artist at heart or anything like that, so drawing a 3D flower and having it visibly come off as that was really interesting to see. I was really proud of that to the point I took screenshots myself of it when I finished because I was really proud of it.''~-P10a
\end{quotation}

\begin{quotation}
``I think it's just that you have this idea in mind and you pretty much just bring it into a world by either making the bear do something or by drawing it. It really brings your thoughts into the real world. I think that that's also a very big component and what makes it meaningful is that you created it, so really, it is yours.''~-P12b
\end{quotation}

With the benefits of agency, the digital handcrafting process became meaningful to participants, and sparked joy in the process of creating for someone else. As such, participants became more willing to invest effort into creation:

\begin{quotation}
``I think the most meaningful part was when I saw it come together as it was loading in for me to send to my partner. That was the part where I realized that I had created every single bit of that message...the drawing and the shirt and that storyline, I created without my own knowledge. That was really meaningful to me because I'm not outwardly a creative person. I found that in this app, I was able to explore those features a little more and find some joy in what I created, even if it wasn't perfect.''~\lei{-P10a}
\end{quotation}




\subsubsection{Treating the craft as a gift encouraged effort}\label{gift}
Many participants treated a crafted \bear{} like a gift or a hand-drawn or handwritten card. In particular, participants drew from the agency they had over handcrafting to treat the craft like a gift.
Since they were given several crafting options, they described the process as similar to spending effort in preparing a gift or a gifting a card to their partner.

\begin{quotation}
``I feel like it's like you want to make a gift. Auggie is like you're trying to create something. Because the bear is like a small board, and you're trying to add image, motions, some actions, musics, voice message, so it's like you're trying to create something... It's like when you create something, you will take more effort to do that... to send something that's very caring for them.''~\lei{-P8b}
\end{quotation}

Participants also described the process of sending their \bear{}s to their partner as a special experience, where blowing on the AR airplane felt like taking a ``moment'' of care:

\begin{quotation}
``I think of when you're sending it off there's a really the intentional blow into your microphone. You send off a little bear, a little airplane. It's like, `Oh yes, it's come all the way to me.' It's a lot different than this is a text message it's going to loop like right across the wireless networks and everything like that. You got a little considerate moment. That's like, `Oh yes, I'm sending off the little thing. Goodbye.'''~-P12a
\end{quotation}




Participant's perceptions of \bear{}s as a gift were further reinforced by their experiences receiving \bear{}s from their partner.
Participants reported feeling a sense of anticipation and gratitude when receiving an \bear{}, similar to feelings they would have when receiving a gift or card in real life. 
For example, one participant talked about the excitement of not knowing what the \bear{} would bring, comparing the process of viewing it to unwrapping a gift:

\begin{quotation}
``...really just seeing the little story unfold. My partner doesn't talk to me before I view it to say, this is what I said or anything. Obviously, it's a surprise when [I watch] it. This also goes into gift-giving, it's almost like unwrapping a present. It's a little bit of suspense and seeing like, oh, what's going to happen here?''~-P12b
\end{quotation}

Participants also described feeling of gratitude when receiving \bear{}s, similar to receiving a gift, due to the effort they perceived was involved in crafting it:

\begin{quotation}
``I feel like there's a lot of similarities [between an Auggie and a real-life gift]. I feel like one of the biggest ones is whenever I receive a gift, I feel like a little bit grateful, that bit of gratitude because you think to yourself like, `Oh, man.' The person who's giving you a gift or whatever, they've put in a lot of work into it.''~-P12a
\end{quotation}

Finally, prior work has found that the motivations of gift-giving include not only social norms such as birthdays, but also altruism (e.g. showing comfort)~\cite{wolfinbarger1990motivations}.
Aligned with this work, we found that participants felt encouraged to invest effort into crafting \bear{}s like gifts for not only planned occasions, such as birthdays, but also day-to-day occasions, such as providing emotional support to their partner: 


\begin{quotation}
``As well as it allows for a fun personalized message, which could be used in that gifting aspect because you could use this incentives to create a personalized Happy Birthday to someone that you might not see all the time or be able to express some emotion a little more directly and more personable than via social media.''~-P10a
\end{quotation}

\begin{quotation}
``Both of us are actually sick at the moment so she stayed home the other day, so that's when I sent that [\bear{}]. I know that that had made her laugh, to see it jumping around and moving so quickly. I know that it was nice for her to hear that I missed her even though she was sick.''~\lei{-P10b}
\end{quotation}


\subsubsection{Crafting stories beyond physical limits encouraged effort}
Participants were also motivated to invest effort due to the ability to create stories beyond their physical limits.
As mentioned in the previous section, the design of \sys{} provides a set of crafting tools that opens up creative possibilities.
Some participants saw the virtual bear character as a representation of themselves that they could equip with new abilities or actions that would be physically difficult for them to achieve in real life. 
With the new capabilities, participants were encouraged to tell imaginary or past stories.
For example, one participant described an \bear{} that they created for their friend, where they animated the bear to go out into the rain when they could not:

\begin{quotation}
``...
my friend just moved from [my city] to another city and it is raining. I [made the] auggie, I [recorded the environment] outside of my window and I made the bear walk around the grass and I [recorded] a video and sent it... I can't go out there if it is raining but the bear can... If I want to jump [on the table], I can't jump on the table but the bear can jump on the table.''~-P8a
\end{quotation}


In addition to physically difficult experiences, participants felt encouraged to invest effort since they could express themselves in new ways that they might not feel comfortable doing in real life.
For example, one participant created an \bear{} in which the bear banged its head against a wall to show that they were cringing after a meeting:
\begin{quotation}
``It could be comic relief, and it's easier to convey the head banging against a wall with the bear instead of me doing that because it will look strange if I show a video of myself doing that.''~-\tc{P2b}
\end{quotation}

Lastly, by blending virtual content with the physical environment, AR offered a sense of embodiment and situatedness, such that participants felt that the virtual content actually exists within the physical world and could interact with the environment in an unconventional way.
Such interactions triggered serendipitous experiences that motivated them to invest effort in order to tell a unique story:

\begin{quotation}
``I went to the kitchen, and we [have a] cast iron, so we don't move it very often. It was just sitting on the stove, and I pointed the camera at it, it takes a second, and the [bear] rendered, and it just rendered it directly standing in the pan and I started laughing. I was like, `I'm just--great, I'm going to send this, this is hilarious, it's just standing in my pan.' Yes, so, I drew a heart, I was laughing, I was like, I swear I'm not cooking the \bear{}...was like my audio note that went with it. It was just funny. Again, it was that real interaction with the environment I think is what made it special.''~-P5a
\end{quotation}

This potential for unexpected and physically impossible interactions with the physical environment (e.g., a bear standing on a pan) can trigger users to invest more effort into creating a whole story and providing context around that interaction, such as adding 3D drawing and voicenotes while expressing their surprise and joy.

\subsubsection{Limits of expressiveness and creativity support discouraged effort.}\label{results_expression}
While the design of \sys{} encouraged effort through its gift-like and storytelling crafting process, some participants reported their unwillingness to invest effort with \sys{} due to limitations in its expressiveness for certain contexts. 
For example, one participant commented that \sys{} is not suitable for serious situations:

\begin{quotation}
``My friend was going through some family stuff the last two weeks. Her dad was in the hospital, so that's more of a serious and emotional topic. It didn't feel very personal or sensitive to send my communications about that with her through Auggie, mainly probably because it feels, although Auggie's an extension of me, its like coming from an AR bear versus me personally''~\lei{-P4a}
\end{quotation}

In this case, crafting with our tool could be inappropriate and undesirable to users, where a different level of personal communication may be necessary.
In addition, our system can only send the virtual bear character in AR, excluding any physical objects from the sender's environment.
This design discouraged some participants from investing effort into crafting as they were unable to accomplish the expressions they wanted with their surrounding objects, which they felt were essential to creating personal stories.
For example, one participant described her frustration with the inability for the bear to interact with their household objects or her cat:

\begin{quotation}
``I would like [to] dance the little bear around my wineglass or have the bear petting my cat but the cat or the wineglass doesn't really go to her [experience]. I'm having the bear do these funny things, but she's not seeing it in my environment.''~-\changed{P4b}
\end{quotation}

Finally, several participants faced eventual creativity blocks in the crafting process. This discouraged effort as participants did not know what to create.
For example, one participant reported having a difficult time coming up with ideas for an \bear{}, heightened by their static physical environment over the course of the study:

\begin{quotation}
``I think my problem was I ran out of ideas and I've also ran out of new scenery...I think if I had new scenery, it'd give me more ideas how to use it, but I'm mostly just sitting in front of my desk every single day so as a result, my \bear{}s, I just haven't gotten any new inspiration.''~\lei{-P1b}
\end{quotation}

\subsection{The effects of digital handcrafting}
We also sought to understand how the effort encouraged through digital handcrafting might affect participants' perceptions of the experience (both as a sender and receiver) and connections with their partner. 
We found that the design of \sys{} had the potential to influence more authentic social connection between partners that felt effortful yet lightweight. 
Participants also reported feelings of social support and presence through handcrafted digital experiences.
We describe these in detail below.

\subsubsection{Personal and authentic connection.}
Our results suggest that effort through digital handcrafting can connect people in more personal and authentic ways, specifically by reflecting the creator's  personal expressions.
In \sys{}, the bear's animations and recorded audio felt especially personal, because the creator's postural and auditory characteristics were embodied in the bear. As previously mentioned, participants often viewed the bear as representative of themselves; thus, they could animate it to express their internal feelings or simulate their actions. One participant compared this process to existing social platforms:

\begin{quotation}
``There is a lot of toxicity found in social media, and this false portrayal of people's lives sometimes. With \sys{}, I think that removes those from the equation, and actually allows you to just be yourself by creating what you want to or creating something for your partner that will be meaningful, without feeling the pressure of society to do something a certain way or make it look a certain way or anything like that. You can just purely have fun with it and enjoy your time using it without those pressures present.... No \bear{} will be like another \bear{}. Whereas there are Snapchat filters where it will make you look like someone that you've never met. With this, you really get to just be yourself and have fun with it.''~-P10a
\end{quotation}

Some also described \sys{}'s potential for inside jokes and comic relief (P2a, P12a), enabling expressions that are comfortably authentic representations of themselves.
For example, one participant created an animation of the bear leaving work and walking out of the lab, as if it was themselves:
\begin{quotation}
``The benefit that I really like is the movement of \bear{}s, which is a personified version of you. It shows your partner or whoever you're sending it to what you're doing. It already feels a little bit more personal [than other communication tools].''~-P2b
\end{quotation}

\subsubsection{Lightweight connection}
We found that participants viewed their communication through digital handcrafting as low-stress and low-urgency due to its resemblance to a letter, greeting card, or postcard (as described in section \ref{gift}). 
Similar to receiving letters and cards in real life, receiving an \bear{} is exciting and meaningful, but does not require an immediate response through \sys{} or other communication means (e.g., text, phone call). 
Aligned with this, 70\% of participants reported in their surveys that they did not expect an immediate response from their partner for the \bear{}s they sent.
Therefore, despite the effort necessary to create \bear{}s, the asynchronicity of the handcrafting experience made it feel \textit{lightweight}, an informal and playful way to convey something interesting or spontaneously show ``I'm thinking of you.'' 

\begin{quotation}
``I guess it's just the fact that I'm thinking about a lot throughout the day and they're very small, not very important thoughts, they're random thoughts or information that I read about online. Just these very small things that maybe are not important enough to save for later to discuss or stuff can be shown through this Auggie since they're a really quick and low commitment.'' -P11a
\end{quotation}

This shows that the feeling of lightweight and low commitment resulted in a casual way of using our system, similar to a conversation starter. 
This approach demonstrates further resemblance to how a postcard is used in communication: effortful in the creation and sending process yet lightweight enough to serve as a catalyst to deeper conversations. 
For P11a, they were able to express random thoughts such as a song that got stuck in their head or a meme they saw online that they wanted to share with their partner, using tools such as \sys{}'s music and customizable outfit features. While participants could have expressed these ideas through other communication means, such as messaging apps, P3b commented on the value of crafting them through \sys{}:

\begin{quotation}
``
[\sys{}] may be more like a postcard. [It's] not quite an instant message. When you sign it, your partner may check it after a while and your message in \sys{} is quite vivid, not like other instant communicate [sic] apps like WeChat or Facebook or something.''~-P3b
\end{quotation}

This aspect of \sys{}'s design consequently lead some participants to feel closer to each other at the end of the study. For example, one participant expressed a greater appreciation between her and her partner because of these lightweight ``conversation starters'' (P2b):

\begin{quotation}
``We’ve been getting closer... I definitely felt like she appreciates our time more or our interactions more and I do too... I think it does add something to our relationship in an interesting sense. We've gotten to talk a lot more because of it.''~-P2b
\end{quotation}




\subsubsection{Feelings of presence}

Participants felt that the situatedness of handcrafting in AR and inclusion of the partner bear character helped enable a sense of presence.
Similar to how participants viewed the bear as a self-representation, participants viewed the partner bear as a representation for their partner.
By superimposing the virtual representations of participants into their partner's physical world, the design of \sys{} helped create a feeling of ``being there'':

\begin{quotation}
``To me, it's the idea of putting some avatar of yourself for the recipient in the scene. You know a good analogy is when you want to be with your friends but one of them is missing and you take a group photo and you hold up your hand and then you're like, `You should be here, your head should be where my hand is.' It's like that. To me, it's sort of that idea. It's like some sense of you should be here, you would enjoy this or like -- yes, that's really what it resonates as for me... you make them dance or move around together, or you try to make them hug or something... that's where it gives me that connection of like my sister and I were sitting together in person somewhere....''~-P5a
\end{quotation}

This highlights how the bear character could be used as a proxy for the recipient, enabling creators to incorporate their partners in the stories they were creating. 
This included scenarios where the bear representations of both people could interact with each together, thus contributing to feelings of togetherness. 
At the same time, this feeling of presence can have an emotional and social impact, which helped participants feel closer:

\begin{quotation}
``One thing [about \sys{}] was since we stay far apart, it is easier to connect. I feel like I'm right there with her, so it's kind of give me the warmth or-- When I miss her, it's like I'm right there with her. It's kind of an intimate or warm feeling that I get from that. That is what I feel like.''~-P14b
\end{quotation}


\section{Discussion}


Our results demonstrate that the design of \sys{} offers a sense of handcraftedness that encourages digital effort, and subsequently has the potential to encourage feelings of personal connection and presence.
At the same time, our findings suggest that effortful communication should be carefully designed, as effort in the wrong contexts could lead it to become meaningless and undesirable.

As prior research has pointed out, effort is widely understood as a desirable concept in interpersonal relationships, as it is interpreted as a sign of mutual affection and care \cite{burke2016facebook}. However, \textit{personal} effort invested in communicating within these relationships is frequently entangled with the concept of \textit{procedural} effort, referring to the work required to operate an interactive system \cite{markopoulos2009design}. This nuance can be lost in the design of communication systems, where \textit{any} effort is systematically perceived as difficult to use, negative, and thus minimized in a system \cite{measuringUX2008}. Message Builder, for example, involves both personal and procedural effort in accessing and writing lengthy text messages, but some users perceived the system as demanding and inconvenient~\cite{kelly2018s}. 

In comparison, our approach attempts to provoke effort in a digital context by making the highly \textit{personal} effort task of handcrafting less \textit{procedurally} effortful. Our findings showed that the experience of using \sys{} was indeed perceived as effortful on a personal level, and, despite investments in time and experimentation, still described as lightweight and approachable on a procedural level. The results suggest that designers can leverage playfulness, flexibility, and asynchronicity to integrate effort into communication for greater social connection.

Based on the concept of procedural and personal effort, we highlight the opportunities and challenges of effortful communication over technology, and discuss design recommendations drawn from our results. We focus our discussions on these two types of effort within social communication systems: on the \textit{procedural} side, making effort approachable by reducing the barrier of interacting with the system; and on the personal side, making effort meaningful through creating unique and personal content and expressions for communication.










\subsection{Designing for approachable effort}
While prior work suggests that procedural effort can be valuable when conveying personal commitment~\cite{kelly2017demanding}, excessive procedural effort can have significant impact on a user's motivation to interact with a system~\cite{kelly2018s,song2021crank}.
Based on our findings, we discuss implications for design that can make effort more \textit{approachable} to people, focused on reducing procedural barriers associated with interacting with a digital communication system.

\subsubsection{Masking the heaviness of effort with playfulness} Excessive procedural effort can have an innate sense of heaviness or undesirability \cite{markopoulos2009design}. We found that the design of \sys{}'s effortful experience, including playful animations and engaging visuals, helped make the procedural aspects of crafting (i.e., the step-by-step of putting the craft together) fun and enjoyable. This demonstrates how playful design can distract people from the potential tedium of required effort with a joyful, more lightweight experience. This aligns with game design research, where designers who aim to change perceptions or teach about conventionally ``heavy'' or ``boring'' topics, like unconscious biases or cybersecurity, face resistance from users. To solve this, researchers have created interventions that use playful content, such as entertaining storylines or cute imagery, to distract from the heaviness of the topic \cite{kaufman2015psychologically,chen2020hacked}. We suggest that future researchers and practitioners exploring effortful communication integrate playfulness alongside effort to reduce the barriers of undesirable procedural work. 
For example, future effortful communication systems might employ similar storylines to generate or deliver the content being communicated (e.g., through whimisical characters or objects), or utilize ``gamification'' attributes such as stickers or reward systems based on effort. At the same time, we highlight the importance of carefully designing such playful features so as not to take away from intrinsic, personal motivations behind effort.

\subsubsection{Lowering the skill required to generate content.} Communication inherently involves the generation of content to be communicated, a process that could involve a lot of skill depending on the content or background of the sender. Moreover, handcrafting that content (e.g., creating it from scratch) can feel more personal~\cite{frizzo2020genuine,fuchs2015handmade}, but require greater expertise (e.g., in drawing or animation).
In our research, we aimed to reduce the barrier of entry to content creation by providing a small and compact set of intuitive creation tools, and found that even people who self-described as non-artistic were able to create experiences that were personal, relatively complex, and aesthetically pleasing. This demonstrates the benefits of designing simplified, more widely accessible tools that can create sophisticated personal content for effortful communication.

Future work might also draw inspiration from past research has that has investigated how to lower the barrier for creating communication content, particularly involving system support. For instance, the ``Lily'' system makes lyrical recommendations to support people in expressing affection~\cite{loveinlyrics2019}. For artistic content, researchers have developed systems that can generate digital outlines to guide novices in drawing~\cite{cheema2012quickdraw}, or correct imperfections in drawings~\cite{iarussi2013drawing}. At the same time, polished content could feel less ``handcrafted'' or personal if the system has a high degree of control over the output compared to the user, replacing the user's efforts~\cite{monroy2011computers, jakesch2019ai}. Future research should further explore the appropriate balance for system support versus user agency that can lower barriers to content creation while retaining authenticity.

\subsubsection{Reducing pressure with asynchronicity.} The asynchronous design of \sys{} helped reduce the pressure to communicate on both the sender's and receiver's side. As suggested by prior work~\cite{kelly2018s}, asynchronicity can ensure that a sender has sufficient time to invest effort in their communication. Compared to a more synchronous exchange, such as a back-and-forth conversation, participants could take the time to carefully craft an experience for their partner. Similarly, on the receiver's end, participants naturally related receiving \bear{}s to receiving a greeting card, i.e., something that one can treasure and keep but not necessarily respond to immediately. 
Consequently, the crafted content can become a precious moment to be treasured, rather than hogging on one's mental capacity.  
Thus, future researchers and practitioners might consider optimizing for asynchronicity, where people have time to craft and reflect on the content of their communication. For example, this could include an asynchronous co-creation process, where people can collaboratively craft content based on their own time and reflect on each others' additions to the craft, or virtual ``mailbox'' spaces that people can join at any time and create content to leave behind for others.

\subsection{Creating meaning in effort} 
In this section, we highlight design implications for heightening meaningful \textit{personal} effort in communication, which, as our results suggest, has the potential to facilitate feelings of connection and presence. In particular, we focus on characteristics that can help frame communication as a \textit{gifting experience}. From our study, participants described finding \bear{}s meaningfully effortful because they felt like gifts. Interestingly, \tc{this is distinct from prior work suggesting that digital gifts lack specific elements that make physical gifts more meaningful in comparison \cite{kwon2017s}}. Building on this work, our results suggest that the design of \sys{} incorporated key attributes and rituals of gifting that can build meaningfulness, such as personalization, ``physical'' wrapping and exchange, and visible effort~\cite{kwon2017s}. We detail three potential design directions along these dimensions.

\subsubsection{Providing channels for personalization} Spence suggests that efforts to personalize a gift and create shared memories can provide a sense of \textit{inalienability}, or the sense that the ``spirit'' of the gift giver remains with the gift even after it's sent, making it unique from simply shared content in communication~\cite{spence2019inalienability}. Our study suggests that a variety of creation channels that act as a scaffold to guide effort~\cite{kelly2018s} can help to achieve personalization.
The variety of channels allows the participants to create personalized communication content for the recipient that feel less like a ``message'' and more like a story through combinations of different elements, such as characters, animations to represent actions, drawings for props, and background music for mood.

We encourage future researchers and designers to consider a range of effortful ``building blocks,'' where people can invest effort into assembling content that is unique and personal for someone else. For example, in the context of social media, Snapchat and Instagram emulate this by providing the function for the users to add in stickers, text, sound, and filters to the photos they share. For text messaging applications, this personalization could happen in not only the content of the message, such as through kinetic typography or font options, but also the application background, message bubbles, and sticker options, and keyboards~\cite{griggio2019customizations,griggio2021mediating}. 
Going beyond communication applications that use 2D content (e.g. images and texts), future work can also explore possible channels of personalizing 3D content (e.g. 3D scenes and characters) for others, such as customizing facial and postural expressions of virtual characters or customizing elements of the virtual environment like the background or surrounding objects.

\subsubsection{Bridging physicality} 
Prior work suggests that physicality can bring meaning to gifting, particularly in the rituals of wrapping and co-located exchange~\cite{kwon2017s,koleva2020designing}. While \sys{} did not involve physical gift wrapping or co-location, we found that the situatedness and physical interactions incorporated in the design helped convey a sense of physicality.
For example, our results suggest the \textit{process} of sending and receiving the content acted as a ``hybrid wrapping,'' or the wrapping of digital gifts in physical material and vice versa~\cite{koleva2020designing}.  Senders took a ``considerate moment'' in blowing wind to send their \bear{} off on an airplane, and receivers landing the airplane and opening the \bear{} felt like they were unwrapping a gift. Future work might consider \tc{designing} similar physical interactions to emulate this effortful wrapping process, such as body gestures (e.g., tossing the content into the air) or facial expressions (e.g., opening it with with a smile).

For the content itself, \bear{}s were played in the receiver's environment, but also included the sender's environment as part of the content through the ``behind-the-scenes'' recordings, which participants enjoyed viewing for additional context (such as in the case of P5a showing the bear on a frying pan). Since the physical environment is fluid and unique for every user, the ability to incorporate any surrounding physical object could be beneficial for encouraging people to craft more personalized and creative experiences that embed aspects of their environment. At the same time, nudging senders to consider crafting the content according to the receiver's environment (e.g., knowing that the receiver usually sits at their desk during the day and would view the content there) could demonstrate more personal effort through the quality of accounting for the receiver's context.

\subsubsection{Explicitly revealing effort} Kwon et al. suggest that people may not value digital gifts when the effort behind them is not evident~\cite{kwon2017s}. Even if a lot of effort was invested, people may not recognize this effort when only the final content is visible, rather than the holistic experience of creating it. Aligned with this, our results suggest that explicitly visualizing effort (e.g., through a ``behind-the-scenes'' screen recording) as opposed to expecting recipients to implicitly recognize effort (e.g., through knowledge about how much effort a system involves) can strengthen the meaningfulness of the communication content and encourage reciprocation.


Researchers and practitioners should thus consider ways to explicitly reveal sender and receiver efforts involved in communication. For example, platforms like Honk\footnote{https://honk.me} use live messaging to make the thoughtful composition of a message and revision process more explicit. \lei{Kelly et al. proposed a concept called Craft Box that uses a video replay of the sender's text composition process to reveal their interactions in the message creation~\cite{kelly2017demanding}.}
Similarly, \sys{} uses a screen recording to show the exact creation process, given the variety of visual elements involved. \lei{While making it easier to understand ``authentic'' effort}, such full recordings may not be desirable for lengthier or more complex effortful processes, as \lei{they may introduce ``higher costs'' for recipients to interpret and acknowledge the effort~\cite{kelly2018designing}}. Senders may also have as privacy concerns with recordings, as they might not feel comfortable exposing the whole creation context to recipients. A less invasive approach could involve aggregate statistics that quantify the effort process, such as the time spent, the number of interactions (i.e., procedural tasks), the number of revisions, and so on to summarize the effort invested. Visualizing traces of effort, such as strokes indicating gestures used on an interface, could provide a more abstract view into the content creation process, while also potentially providing a sense of presence~\cite{monastero2018traces}.


\subsection{Limitations}
Our results demonstrate that \sys{} encourages effortful handcrafting of digital artifacts, which subsequently has the potential to support meaningful interactions with close others. 
However, our work is not without limitations. 

First, while deploying \sys{} in participants' everyday lives enabled evaluating the system \textit{in situ}, the nature of a field study limits our ability to control for external variables, such as other communication channels or effortful practices the participants might have engaged in during the study. This can subsequently affect our ability to isolate the specific effects of our system. Future work should consider more controlled experiments or the inclusion of comparison groups to better measure the impact of effort and digital handcrafting on social connection and interpersonal relationships.

Second, \changed{requesting} participants to create at least one \bear{} a day allowed us to understand the perceptions of the system under frequent usage. However, this requirement may have limited the scenarios in which effort was meaningful, as participants noted that this frequency did not necessarily align with how they believed they would normally use the system. \os{This may explain the ``creativity blocks'' discussed in Section \ref{results_expression} that some participants experienced during the crafting process, where participants described their everyday environment as unchanging}. Future work should explore how the regularity of such interactions and greater variety of social situations might affect people's perception and desire to engage in effort, such as through longer studies across several months. \os{Additionally, while we did not observe specific novelty effects during the study, it is possible that it played a role, and a longer study would be beneficial to understand how people would use systems like \sys{} in steady state.}

Finally, participants primarily participated with close study partners (i.e., friends, family, significant others). While we still gleaned valuable and diverse insights from their usage of \sys{}, participants in different relationship types may have different perceptions of the application. For example, early stage romantic couples may inherently feel motivated to invest more effort more frequently, as part of strengthening their relationship. Future researchers should thus consider investigating the usage and value of effortful communication for different relationship types.

\section{Conclusion}
We introduced \sys{}, an iOS app for engaging in effortful communication through digitally handcrafted AR experiences. Through a two-week long field deployment with 30 participants (15 pairs), we explored how to encourage meaningful, personal effort in digital communication systems, and its subsequent effects on relationships. 
Our research revealed that the design of \sys{} inspired participants to invest effort in crafting personalized experiences by evoking a sense of agency that enabled them to craft gift-like personal stories beyond physical limits.
These immersive, digitally handcrafted experiences subsequently had the potential to enable authentic, lightweight connection and feelings of presence between partners.
At the same time, we found that limited expressiveness and creativity support for different situations  discouraged people in engaging in effortful communication.
We yield insights and recommendations that provide a starting point for researchers and practitioners to further explore meaningful personal effort in digital communication.

\bibliographystyle{ACM-Reference-Format}
\bibliography{references}


\begin{thebibliography}{62}


\ifx \showCODEN    \undefined \def \showCODEN     #1{\unskip}     \fi
\ifx \showDOI      \undefined \def \showDOI       #1{#1}\fi
\ifx \showISBNx    \undefined \def \showISBNx     #1{\unskip}     \fi
\ifx \showISBNxiii \undefined \def \showISBNxiii  #1{\unskip}     \fi
\ifx \showISSN     \undefined \def \showISSN      #1{\unskip}     \fi
\ifx \showLCCN     \undefined \def \showLCCN      #1{\unskip}     \fi
\ifx \shownote     \undefined \def \shownote      #1{#1}          \fi
\ifx \showarticletitle \undefined \def \showarticletitle #1{#1}   \fi
\ifx \showURL      \undefined \def \showURL       {\relax}        \fi
\providecommand\bibfield[2]{#2}
\providecommand\bibinfo[2]{#2}
\providecommand\natexlab[1]{#1}
\providecommand\showeprint[2][]{arXiv:#2}

\bibitem[\protect\citeauthoryear{Bai, Blackwell, and Coulouris}{Bai
  et~al\mbox{.}}{2015}]%
        {bai2015exploring}
\bibfield{author}{\bibinfo{person}{Zhen Bai}, \bibinfo{person}{Alan~F
  Blackwell}, {and} \bibinfo{person}{George Coulouris}.}
  \bibinfo{year}{2015}\natexlab{}.
\newblock \showarticletitle{Exploring expressive augmented reality: The FingAR
  puppet system for social pretend play}. In
  \bibinfo{booktitle}{\emph{Proceedings of the 33rd Annual ACM Conference on
  Human Factors in Computing Systems}}. \bibinfo{pages}{1035--1044}.
\newblock


\bibitem[\protect\citeauthoryear{Bao, Herlocker, and Dietterich}{Bao
  et~al\mbox{.}}{2006}]%
        {bao2006fewerclicks}
\bibfield{author}{\bibinfo{person}{Xinlong Bao}, \bibinfo{person}{Jonathan~L.
  Herlocker}, {and} \bibinfo{person}{Thomas~G. Dietterich}.}
  \bibinfo{year}{2006}\natexlab{}.
\newblock \showarticletitle{Fewer Clicks and Less Frustration: Reducing the
  Cost of Reaching the Right Folder}. In \bibinfo{booktitle}{\emph{Proceedings
  of the 11th International Conference on Intelligent User Interfaces}}
  (Sydney, Australia) \emph{(\bibinfo{series}{IUI '06})}.
  \bibinfo{publisher}{Association for Computing Machinery},
  \bibinfo{address}{New York, NY, USA}, \bibinfo{pages}{178–185}.
\newblock
\showISBNx{1595932879}
\urldef\tempurl%
\url{https://doi.org/10.1145/1111449.1111490}
\showDOI{\tempurl}


\bibitem[\protect\citeauthoryear{Biocca, Harms, and Burgoon}{Biocca
  et~al\mbox{.}}{2003}]%
        {biocca2003toward}
\bibfield{author}{\bibinfo{person}{Frank Biocca}, \bibinfo{person}{Chad Harms},
  {and} \bibinfo{person}{Judee~K Burgoon}.} \bibinfo{year}{2003}\natexlab{}.
\newblock \showarticletitle{Toward a more robust theory and measure of social
  presence: Review and suggested criteria}.
\newblock \bibinfo{journal}{\emph{Presence: Teleoperators \& virtual
  environments}} \bibinfo{volume}{12}, \bibinfo{number}{5}
  (\bibinfo{year}{2003}), \bibinfo{pages}{456--480}.
\newblock


\bibitem[\protect\citeauthoryear{Brunell, Kernis, Goldman, Heppner, Davis,
  Cascio, and Webster}{Brunell et~al\mbox{.}}{2010}]%
        {brunell2010dispositional}
\bibfield{author}{\bibinfo{person}{Amy~B Brunell}, \bibinfo{person}{Michael~H
  Kernis}, \bibinfo{person}{Brian~M Goldman}, \bibinfo{person}{Whitney
  Heppner}, \bibinfo{person}{Patricia Davis}, \bibinfo{person}{Edward~V
  Cascio}, {and} \bibinfo{person}{Gregory~D Webster}.}
  \bibinfo{year}{2010}\natexlab{}.
\newblock \showarticletitle{Dispositional authenticity and romantic
  relationship functioning}.
\newblock \bibinfo{journal}{\emph{Personality and Individual Differences}}
  \bibinfo{volume}{48}, \bibinfo{number}{8} (\bibinfo{year}{2010}),
  \bibinfo{pages}{900--905}.
\newblock


\bibitem[\protect\citeauthoryear{Burke and Kraut}{Burke and Kraut}{2016}]%
        {burke2016facebook}
\bibfield{author}{\bibinfo{person}{Moira Burke} {and}
  \bibinfo{person}{Robert~E. Kraut}.} \bibinfo{year}{2016}\natexlab{}.
\newblock \showarticletitle{{The Relationship Between Facebook Use and
  Well-Being Depends on Communication Type and Tie Strength}}.
\newblock \bibinfo{journal}{\emph{Journal of Computer-Mediated Communication}}
  \bibinfo{volume}{21}, \bibinfo{number}{4} (\bibinfo{date}{05}
  \bibinfo{year}{2016}), \bibinfo{pages}{265--281}.
\newblock
\showISSN{1083-6101}
\urldef\tempurl%
\url{https://doi.org/10.1111/jcc4.12162}
\showDOI{\tempurl}
\showeprint{https://academic.oup.com/jcmc/article-pdf/21/4/265/22316334/jjcmcom0265.pdf}


\bibitem[\protect\citeauthoryear{Canary and Stafford}{Canary and
  Stafford}{1992}]%
        {canary1992relational}
\bibfield{author}{\bibinfo{person}{Daniel~J Canary} {and}
  \bibinfo{person}{Laura Stafford}.} \bibinfo{year}{1992}\natexlab{}.
\newblock \showarticletitle{Relational maintenance strategies and equity in
  marriage}.
\newblock \bibinfo{journal}{\emph{Communications Monographs}}
  \bibinfo{volume}{59}, \bibinfo{number}{3} (\bibinfo{year}{1992}),
  \bibinfo{pages}{243--267}.
\newblock


\bibitem[\protect\citeauthoryear{Canary and Yum}{Canary and Yum}{2015}]%
        {canary2015relationship}
\bibfield{author}{\bibinfo{person}{Daniel~J Canary} {and}
  \bibinfo{person}{Young-Ok Yum}.} \bibinfo{year}{2015}\natexlab{}.
\newblock \showarticletitle{Relationship maintenance strategies}.
\newblock \bibinfo{journal}{\emph{The international encyclopedia of
  interpersonal communication}} (\bibinfo{year}{2015}), \bibinfo{pages}{1--9}.
\newblock


\bibitem[\protect\citeauthoryear{Cheema, Gulwani, and LaViola}{Cheema
  et~al\mbox{.}}{2012}]%
        {cheema2012quickdraw}
\bibfield{author}{\bibinfo{person}{Salman Cheema}, \bibinfo{person}{Sumit
  Gulwani}, {and} \bibinfo{person}{Joseph LaViola}.}
  \bibinfo{year}{2012}\natexlab{}.
\newblock \showarticletitle{QuickDraw: improving drawing experience for
  geometric diagrams}. In \bibinfo{booktitle}{\emph{Proceedings of the SIGCHI
  Conference on Human Factors in Computing Systems}}.
  \bibinfo{pages}{1037--1064}.
\newblock


\bibitem[\protect\citeauthoryear{Chen, Stewart, Bai, Chen, Dabbish, and
  Hammer}{Chen et~al\mbox{.}}{2020}]%
        {chen2020hacked}
\bibfield{author}{\bibinfo{person}{Tianying Chen}, \bibinfo{person}{Margot
  Stewart}, \bibinfo{person}{Zhiyu Bai}, \bibinfo{person}{Eileen Chen},
  \bibinfo{person}{Laura Dabbish}, {and} \bibinfo{person}{Jessica Hammer}.}
  \bibinfo{year}{2020}\natexlab{}.
\newblock \showarticletitle{Hacked Time: Design and Evaluation of a
  Self-Efficacy Based Cybersecurity Game}. In
  \bibinfo{booktitle}{\emph{Proceedings of the 2020 ACM Designing Interactive
  Systems Conference}}. \bibinfo{pages}{1737--1749}.
\newblock


\bibitem[\protect\citeauthoryear{Croom}{Croom}{2015}]%
        {croom2015music}
\bibfield{author}{\bibinfo{person}{Adam~M Croom}.}
  \bibinfo{year}{2015}\natexlab{}.
\newblock \showarticletitle{Music practice and participation for psychological
  well-being: A review of how music influences positive emotion, engagement,
  relationships, meaning, and accomplishment}.
\newblock \bibinfo{journal}{\emph{Musicae Scientiae}} \bibinfo{volume}{19},
  \bibinfo{number}{1} (\bibinfo{year}{2015}), \bibinfo{pages}{44--64}.
\newblock


\bibitem[\protect\citeauthoryear{Dagan, C{\'a}rdenas~Gasca, Robinson, Noriega,
  Tham, Vaish, and Monroy-Hern{\'a}ndez}{Dagan et~al\mbox{.}}{2022}]%
        {dagan2022project}
\bibfield{author}{\bibinfo{person}{Ella Dagan}, \bibinfo{person}{Ana~Mar{\'\i}a
  C{\'a}rdenas~Gasca}, \bibinfo{person}{Ava Robinson}, \bibinfo{person}{Anwar
  Noriega}, \bibinfo{person}{Yu~Jiang Tham}, \bibinfo{person}{Rajan Vaish},
  {and} \bibinfo{person}{Andr{\'e}s Monroy-Hern{\'a}ndez}.}
  \bibinfo{year}{2022}\natexlab{}.
\newblock \showarticletitle{Project IRL: Playful Co-Located Interactions with
  Mobile Augmented Reality}.
\newblock \bibinfo{journal}{\emph{Proceedings of the ACM on Human-Computer
  Interaction}} \bibinfo{volume}{6}, \bibinfo{number}{CSCW1}
  (\bibinfo{year}{2022}), \bibinfo{pages}{1--27}.
\newblock


\bibitem[\protect\citeauthoryear{Damala, Cubaud, Bationo, Houlier, and
  Marchal}{Damala et~al\mbox{.}}{2008}]%
        {damala2008bridging}
\bibfield{author}{\bibinfo{person}{Areti Damala}, \bibinfo{person}{Pierre
  Cubaud}, \bibinfo{person}{Anne Bationo}, \bibinfo{person}{Pascal Houlier},
  {and} \bibinfo{person}{Isabelle Marchal}.} \bibinfo{year}{2008}\natexlab{}.
\newblock \showarticletitle{Bridging the gap between the digital and the
  physical: design and evaluation of a mobile augmented reality guide for the
  museum visit}. In \bibinfo{booktitle}{\emph{Proceedings of the 3rd
  international conference on Digital Interactive Media in Entertainment and
  Arts}}. \bibinfo{pages}{120--127}.
\newblock


\bibitem[\protect\citeauthoryear{Doub\'{e} and Beh}{Doub\'{e} and Beh}{2012}]%
        {doube2012autocomplete}
\bibfield{author}{\bibinfo{person}{Wendy Doub\'{e}} {and}
  \bibinfo{person}{Jeanie Beh}.} \bibinfo{year}{2012}\natexlab{}.
\newblock \showarticletitle{Typing over Autocomplete: Cognitive Load in Website
  Use by Older Adults}. In \bibinfo{booktitle}{\emph{Proceedings of the 24th
  Australian Computer-Human Interaction Conference}} (Melbourne, Australia)
  \emph{(\bibinfo{series}{OzCHI '12})}. \bibinfo{publisher}{Association for
  Computing Machinery}, \bibinfo{address}{New York, NY, USA},
  \bibinfo{pages}{97–106}.
\newblock
\showISBNx{9781450314381}
\urldef\tempurl%
\url{https://doi.org/10.1145/2414536.2414553}
\showDOI{\tempurl}


\bibitem[\protect\citeauthoryear{Fehnert and Kosagowsky}{Fehnert and
  Kosagowsky}{2008}]%
        {measuringUX2008}
\bibfield{author}{\bibinfo{person}{Ben Fehnert} {and} \bibinfo{person}{Alessia
  Kosagowsky}.} \bibinfo{year}{2008}\natexlab{}.
\newblock \showarticletitle{Measuring User Experience: Complementing
  Qualitative and Quantitative Assessment}. In
  \bibinfo{booktitle}{\emph{Proceedings of the 10th International Conference on
  Human Computer Interaction with Mobile Devices and Services}} (Amsterdam, The
  Netherlands) \emph{(\bibinfo{series}{MobileHCI '08})}.
  \bibinfo{publisher}{Association for Computing Machinery},
  \bibinfo{address}{New York, NY, USA}, \bibinfo{pages}{383–386}.
\newblock
\showISBNx{9781595939524}
\urldef\tempurl%
\url{https://doi.org/10.1145/1409240.1409294}
\showDOI{\tempurl}


\bibitem[\protect\citeauthoryear{Fleishman}{Fleishman}{2018}]%
        {fleishman_2018}
\bibfield{author}{\bibinfo{person}{Glenn Fleishman}.}
  \bibinfo{year}{2018}\natexlab{}.
\newblock \bibinfo{title}{How Facebook Devalued The Birthday}.
\newblock
\newblock
\urldef\tempurl%
\url{https://www.fastcompany.com/40550725/how-facebook-devalued-the-birthday}
\showURL{%
\tempurl}


\bibitem[\protect\citeauthoryear{Frizzo, Dias, Duarte, Rodrigues, and
  Prado}{Frizzo et~al\mbox{.}}{2020}]%
        {frizzo2020genuine}
\bibfield{author}{\bibinfo{person}{Francielle Frizzo}, \bibinfo{person}{Helison
  Bertoli~Alves Dias}, \bibinfo{person}{Nayara~Pereira Duarte},
  \bibinfo{person}{Denise~Gabriela Rodrigues}, {and} \bibinfo{person}{Paulo
  Henrique~Muller Prado}.} \bibinfo{year}{2020}\natexlab{}.
\newblock \showarticletitle{The Genuine Handmade: How the Production Method
  Influences Consumers’ Behavioral Intentions through Naturalness and
  Authenticity}.
\newblock \bibinfo{journal}{\emph{Journal of Food Products Marketing}}
  \bibinfo{volume}{26}, \bibinfo{number}{4} (\bibinfo{year}{2020}),
  \bibinfo{pages}{279--296}.
\newblock


\bibitem[\protect\citeauthoryear{Fuchs, Schreier, and Van~Osselaer}{Fuchs
  et~al\mbox{.}}{2015}]%
        {fuchs2015handmade}
\bibfield{author}{\bibinfo{person}{Christoph Fuchs}, \bibinfo{person}{Martin
  Schreier}, {and} \bibinfo{person}{Stijn~MJ Van~Osselaer}.}
  \bibinfo{year}{2015}\natexlab{}.
\newblock \showarticletitle{The handmade effect: What's love got to do with
  it?}
\newblock \bibinfo{journal}{\emph{Journal of marketing}} \bibinfo{volume}{79},
  \bibinfo{number}{2} (\bibinfo{year}{2015}), \bibinfo{pages}{98--110}.
\newblock


\bibitem[\protect\citeauthoryear{Gloria}{Gloria}{2020}]%
        {lessonsinloneliness}
\bibfield{author}{\bibinfo{person}{Kristine Gloria}.}
  \bibinfo{year}{2020}\natexlab{}.
\newblock \bibinfo{booktitle}{\emph{Lessons in Loneliness}}.
\newblock \bibinfo{type}{{T}echnical {R}eport}. \bibinfo{institution}{Facebook
  and Aspen Institute}. \bibinfo{pages}{28} pages.
\newblock
\urldef\tempurl%
\url{https://www.aspeninstitute.org/blog-posts/future-of-loneliness-social-connection-technology/}
\showURL{%
\tempurl}


\bibitem[\protect\citeauthoryear{Griggio, Sato, Mackay, and Yatani}{Griggio
  et~al\mbox{.}}{2021a}]%
        {dearboard2021}
\bibfield{author}{\bibinfo{person}{Carla Griggio}, \bibinfo{person}{Arissa
  Sato}, \bibinfo{person}{Wendy Mackay}, {and} \bibinfo{person}{Koji Yatani}.}
  \bibinfo{year}{2021}\natexlab{a}.
\newblock \showarticletitle{Mediating Intimacy with DearBoard: a
  Co-Customizable Keyboard for Everyday Messaging}.
\newblock
\urldef\tempurl%
\url{https://doi.org/10.1145/3411764.3445757}
\showDOI{\tempurl}


\bibitem[\protect\citeauthoryear{Griggio, Mcgrenere, and Mackay}{Griggio
  et~al\mbox{.}}{2019}]%
        {griggio2019customizations}
\bibfield{author}{\bibinfo{person}{Carla~F Griggio}, \bibinfo{person}{Joanna
  Mcgrenere}, {and} \bibinfo{person}{Wendy~E Mackay}.}
  \bibinfo{year}{2019}\natexlab{}.
\newblock \showarticletitle{Customizations and expression breakdowns in
  ecosystems of communication apps}.
\newblock \bibinfo{journal}{\emph{Proceedings of the ACM on Human-Computer
  Interaction}} \bibinfo{volume}{3}, \bibinfo{number}{CSCW}
  (\bibinfo{year}{2019}), \bibinfo{pages}{1--26}.
\newblock


\bibitem[\protect\citeauthoryear{Griggio, Sato, Mackay, and Yatani}{Griggio
  et~al\mbox{.}}{2021b}]%
        {griggio2021mediating}
\bibfield{author}{\bibinfo{person}{Carla~F Griggio}, \bibinfo{person}{Arissa~J
  Sato}, \bibinfo{person}{Wendy~E Mackay}, {and} \bibinfo{person}{Koji
  Yatani}.} \bibinfo{year}{2021}\natexlab{b}.
\newblock \showarticletitle{Mediating Intimacy with DearBoard: a
  Co-Customizable Keyboard for Everyday Messaging}. In
  \bibinfo{booktitle}{\emph{Proceedings of the 2021 CHI Conference on Human
  Factors in Computing Systems}}. \bibinfo{pages}{1--16}.
\newblock


\bibitem[\protect\citeauthoryear{Guo, Canberk, Murphy, Monroy-Hern{\'a}ndez,
  and Vaish}{Guo et~al\mbox{.}}{2019}]%
        {guo2019blocks}
\bibfield{author}{\bibinfo{person}{Anhong Guo}, \bibinfo{person}{Ilter
  Canberk}, \bibinfo{person}{Hannah Murphy}, \bibinfo{person}{Andr{\'e}s
  Monroy-Hern{\'a}ndez}, {and} \bibinfo{person}{Rajan Vaish}.}
  \bibinfo{year}{2019}\natexlab{}.
\newblock \showarticletitle{Blocks: Collaborative and persistent augmented
  reality experiences}.
\newblock \bibinfo{journal}{\emph{Proceedings of the ACM on Interactive,
  Mobile, Wearable and Ubiquitous Technologies}} \bibinfo{volume}{3},
  \bibinfo{number}{3} (\bibinfo{year}{2019}), \bibinfo{pages}{1--24}.
\newblock


\bibitem[\protect\citeauthoryear{Hallinan}{Hallinan}{2018}]%
        {hallinan2018like}
\bibfield{author}{\bibinfo{person}{Blake Hallinan}.}
  \bibinfo{year}{2018}\natexlab{}.
\newblock \showarticletitle{Like: The Informatization of Affect}. In
  \bibinfo{booktitle}{\emph{Companion of the 2018 ACM Conference on Computer
  Supported Cooperative Work and Social Computing}} (Jersey City, NJ, USA)
  \emph{(\bibinfo{series}{CSCW '18})}. \bibinfo{publisher}{Association for
  Computing Machinery}, \bibinfo{address}{New York, NY, USA},
  \bibinfo{pages}{73–76}.
\newblock
\showISBNx{9781450360180}
\urldef\tempurl%
\url{https://doi.org/10.1145/3272973.3272977}
\showDOI{\tempurl}


\bibitem[\protect\citeauthoryear{Higuera-Trujillo, Maldonado, and
  Mill{\'a}n}{Higuera-Trujillo et~al\mbox{.}}{2017}]%
        {higuera2017psychological}
\bibfield{author}{\bibinfo{person}{Juan~Luis Higuera-Trujillo},
  \bibinfo{person}{Juan L{\'o}pez-Tarruella Maldonado}, {and}
  \bibinfo{person}{Carmen~Llinares Mill{\'a}n}.}
  \bibinfo{year}{2017}\natexlab{}.
\newblock \showarticletitle{Psychological and physiological human responses to
  simulated and real environments: A comparison between Photographs, 360
  Panoramas, and Virtual Reality}.
\newblock \bibinfo{journal}{\emph{Applied ergonomics}}  \bibinfo{volume}{65}
  (\bibinfo{year}{2017}), \bibinfo{pages}{398--409}.
\newblock


\bibitem[\protect\citeauthoryear{Hsieh and Tseng}{Hsieh and Tseng}{2017}]%
        {hsieh2017playfulness}
\bibfield{author}{\bibinfo{person}{Sara~H Hsieh} {and} \bibinfo{person}{Timmy~H
  Tseng}.} \bibinfo{year}{2017}\natexlab{}.
\newblock \showarticletitle{Playfulness in mobile instant messaging: Examining
  the influence of emoticons and text messaging on social interaction}.
\newblock \bibinfo{journal}{\emph{Computers in Human Behavior}}
  \bibinfo{volume}{69} (\bibinfo{year}{2017}), \bibinfo{pages}{405--414}.
\newblock


\bibitem[\protect\citeauthoryear{Iarussi, Bousseau, and Tsandilas}{Iarussi
  et~al\mbox{.}}{2013}]%
        {iarussi2013drawing}
\bibfield{author}{\bibinfo{person}{Emmanuel Iarussi}, \bibinfo{person}{Adrien
  Bousseau}, {and} \bibinfo{person}{Theophanis Tsandilas}.}
  \bibinfo{year}{2013}\natexlab{}.
\newblock \showarticletitle{The drawing assistant: Automated drawing guidance
  and feedback from photographs}. In \bibinfo{booktitle}{\emph{ACM Symposium on
  User Interface Software and Technology (UIST)}}. ACM.
\newblock


\bibitem[\protect\citeauthoryear{Jakesch, French, Ma, Hancock, and
  Naaman}{Jakesch et~al\mbox{.}}{2019}]%
        {jakesch2019ai}
\bibfield{author}{\bibinfo{person}{Maurice Jakesch}, \bibinfo{person}{Megan
  French}, \bibinfo{person}{Xiao Ma}, \bibinfo{person}{Jeffrey~T Hancock},
  {and} \bibinfo{person}{Mor Naaman}.} \bibinfo{year}{2019}\natexlab{}.
\newblock \showarticletitle{AI-mediated communication: How the perception that
  profile text was written by AI affects trustworthiness}. In
  \bibinfo{booktitle}{\emph{Proceedings of the 2019 CHI Conference on Human
  Factors in Computing Systems}}. \bibinfo{pages}{1--13}.
\newblock


\bibitem[\protect\citeauthoryear{Kaufman and Flanagan}{Kaufman and
  Flanagan}{2015}]%
        {kaufman2015psychologically}
\bibfield{author}{\bibinfo{person}{Geoff Kaufman} {and} \bibinfo{person}{Mary
  Flanagan}.} \bibinfo{year}{2015}\natexlab{}.
\newblock \showarticletitle{A psychologically “embedded” approach to
  designing games for prosocial causes}.
\newblock \bibinfo{journal}{\emph{Cyberpsychology: Journal of Psychosocial
  Research on Cyberspace}} \bibinfo{volume}{9}, \bibinfo{number}{3}
  (\bibinfo{year}{2015}).
\newblock


\bibitem[\protect\citeauthoryear{Kelly, Gooch, Patil, and Watts}{Kelly
  et~al\mbox{.}}{2017}]%
        {kelly2017demanding}
\bibfield{author}{\bibinfo{person}{Ryan Kelly}, \bibinfo{person}{Daniel Gooch},
  \bibinfo{person}{Bhagyashree Patil}, {and} \bibinfo{person}{Leon Watts}.}
  \bibinfo{year}{2017}\natexlab{}.
\newblock \showarticletitle{Demanding by design: Supporting effortful
  communication practices in close personal relationships}. In
  \bibinfo{booktitle}{\emph{Proceedings of the 2017 ACM Conference on Computer
  Supported Cooperative Work and Social Computing}}. \bibinfo{pages}{70--83}.
\newblock


\bibitem[\protect\citeauthoryear{Kelly, Gooch, and Watts}{Kelly
  et~al\mbox{.}}{2018a}]%
        {kelly2018designing}
\bibfield{author}{\bibinfo{person}{Ryan Kelly}, \bibinfo{person}{Daniel Gooch},
  {and} \bibinfo{person}{Leon Watts}.} \bibinfo{year}{2018}\natexlab{a}.
\newblock \showarticletitle{Designing for reflection on sender effort in close
  personal communication}. In \bibinfo{booktitle}{\emph{Proceedings of the 30th
  Australian Conference on Computer-Human Interaction}}.
  \bibinfo{pages}{314--325}.
\newblock


\bibitem[\protect\citeauthoryear{Kelly, Gooch, and Watts}{Kelly
  et~al\mbox{.}}{2018b}]%
        {kelly2018s}
\bibfield{author}{\bibinfo{person}{Ryan Kelly}, \bibinfo{person}{Daniel Gooch},
  {and} \bibinfo{person}{Leon Watts}.} \bibinfo{year}{2018}\natexlab{b}.
\newblock \showarticletitle{'It's More Like a Letter' An Exploration of
  Mediated Conversational Effort in Message Builder}.
\newblock \bibinfo{journal}{\emph{Proceedings of the ACM on Human-Computer
  Interaction}} \bibinfo{volume}{2}, \bibinfo{number}{CSCW}
  (\bibinfo{year}{2018}), \bibinfo{pages}{1--23}.
\newblock


\bibitem[\protect\citeauthoryear{Kim, Lee, Peng, and Ma}{Kim
  et~al\mbox{.}}{2019}]%
        {loveinlyrics2019}
\bibfield{author}{\bibinfo{person}{Taewook Kim}, \bibinfo{person}{Jung Lee},
  \bibinfo{person}{Zhenhui Peng}, {and} \bibinfo{person}{Xiaojuan Ma}.}
  \bibinfo{year}{2019}\natexlab{}.
\newblock \showarticletitle{Love in Lyrics: An Exploration of Supporting
  Textual Manifestation of Affection in Social Messaging}.
\newblock \bibinfo{journal}{\emph{Proceedings of the ACM on Human-Computer
  Interaction}}  \bibinfo{volume}{3} (\bibinfo{date}{11} \bibinfo{year}{2019}),
  \bibinfo{pages}{1--27}.
\newblock
\urldef\tempurl%
\url{https://doi.org/10.1145/3359181}
\showDOI{\tempurl}


\bibitem[\protect\citeauthoryear{King and Forlizzi}{King and Forlizzi}{2007}]%
        {king2007slow}
\bibfield{author}{\bibinfo{person}{Simon King} {and} \bibinfo{person}{Jodi
  Forlizzi}.} \bibinfo{year}{2007}\natexlab{}.
\newblock \showarticletitle{Slow messaging: intimate communication for couples
  living at a distance}. In \bibinfo{booktitle}{\emph{Proceedings of the 2007
  conference on Designing pleasurable products and interfaces}}.
  \bibinfo{pages}{451--454}.
\newblock


\bibitem[\protect\citeauthoryear{Klein}{Klein}{2015}]%
        {tobiasdigitalcraftsmanship}
\bibfield{author}{\bibinfo{person}{Tobias Klein}.}
  \bibinfo{year}{2015}\natexlab{}.
\newblock \showarticletitle{Digital Craftsmanship}. \bibinfo{pages}{643--654}.
\newblock
\showISBNx{978-3-319-20897-8}
\urldef\tempurl%
\url{https://doi.org/10.1007/978-3-319-20898-5_61}
\showDOI{\tempurl}


\bibitem[\protect\citeauthoryear{Koleva, Spence, Benford, Kwon,
  Schn{\"a}delbach, Thorn, Preston, Hazzard, Greenhalgh, Adams,
  et~al\mbox{.}}{Koleva et~al\mbox{.}}{2020}]%
        {koleva2020designing}
\bibfield{author}{\bibinfo{person}{Boriana Koleva}, \bibinfo{person}{Jocelyn
  Spence}, \bibinfo{person}{Steve Benford}, \bibinfo{person}{Hyosun Kwon},
  \bibinfo{person}{Holger Schn{\"a}delbach}, \bibinfo{person}{Emily Thorn},
  \bibinfo{person}{William Preston}, \bibinfo{person}{Adrian Hazzard},
  \bibinfo{person}{Chris Greenhalgh}, \bibinfo{person}{Matt Adams},
  {et~al\mbox{.}}} \bibinfo{year}{2020}\natexlab{}.
\newblock \showarticletitle{Designing hybrid gifts}.
\newblock \bibinfo{journal}{\emph{ACM Transactions on Computer-Human
  Interaction (TOCHI)}} \bibinfo{volume}{27}, \bibinfo{number}{5}
  (\bibinfo{year}{2020}), \bibinfo{pages}{1--33}.
\newblock


\bibitem[\protect\citeauthoryear{Kroupi, Hanhart, Lee, Rerabek, and
  Ebrahimi}{Kroupi et~al\mbox{.}}{2014}]%
        {kroupi2014predicting}
\bibfield{author}{\bibinfo{person}{Eleni Kroupi}, \bibinfo{person}{Philippe
  Hanhart}, \bibinfo{person}{Jong-Seok Lee}, \bibinfo{person}{Martin Rerabek},
  {and} \bibinfo{person}{Touradj Ebrahimi}.} \bibinfo{year}{2014}\natexlab{}.
\newblock \showarticletitle{Predicting subjective sensation of reality during
  multimedia consumption based on EEG and peripheral physiological signals}. In
  \bibinfo{booktitle}{\emph{2014 IEEE International Conference on Multimedia
  and Expo (ICME)}}. IEEE, \bibinfo{pages}{1--6}.
\newblock


\bibitem[\protect\citeauthoryear{Kwon, Koleva, Schn{\"a}delbach, and
  Benford}{Kwon et~al\mbox{.}}{2017}]%
        {kwon2017s}
\bibfield{author}{\bibinfo{person}{Hyosun Kwon}, \bibinfo{person}{Boriana
  Koleva}, \bibinfo{person}{Holger Schn{\"a}delbach}, {and}
  \bibinfo{person}{Steve Benford}.} \bibinfo{year}{2017}\natexlab{}.
\newblock \showarticletitle{" It's Not Yet A Gift" Understanding Digital
  Gifting}. In \bibinfo{booktitle}{\emph{Proceedings of the 2017 ACM Conference
  on Computer Supported Cooperative Work and Social Computing}}.
  \bibinfo{pages}{2372--2384}.
\newblock


\bibitem[\protect\citeauthoryear{Leiva, Nguyen, Kazi, and Asente}{Leiva
  et~al\mbox{.}}{2020}]%
        {leiva2020pronto}
\bibfield{author}{\bibinfo{person}{Germ{\'a}n Leiva}, \bibinfo{person}{Cuong
  Nguyen}, \bibinfo{person}{Rubaiat~Habib Kazi}, {and} \bibinfo{person}{Paul
  Asente}.} \bibinfo{year}{2020}\natexlab{}.
\newblock \showarticletitle{Pronto: Rapid augmented reality video prototyping
  using sketches and enaction}. In \bibinfo{booktitle}{\emph{Proceedings of the
  2020 CHI Conference on Human Factors in Computing Systems}}.
  \bibinfo{pages}{1--13}.
\newblock


\bibitem[\protect\citeauthoryear{Lindley, Harper, and Sellen}{Lindley
  et~al\mbox{.}}{2009}]%
        {lindley2009desiring}
\bibfield{author}{\bibinfo{person}{Si{\^a}n~E Lindley},
  \bibinfo{person}{Richard Harper}, {and} \bibinfo{person}{Abigail Sellen}.}
  \bibinfo{year}{2009}\natexlab{}.
\newblock \showarticletitle{Desiring to be in touch in a changing
  communications landscape: attitudes of older adults}. In
  \bibinfo{booktitle}{\emph{Proceedings of the SIGCHI Conference on Human
  Factors in Computing Systems}}. \bibinfo{pages}{1693--1702}.
\newblock


\bibitem[\protect\citeauthoryear{Litt, Zhao, Kraut, and Burke}{Litt
  et~al\mbox{.}}{2020}]%
        {litt2020meaningful}
\bibfield{author}{\bibinfo{person}{Eden Litt}, \bibinfo{person}{Siyan Zhao},
  \bibinfo{person}{Robert Kraut}, {and} \bibinfo{person}{Moira Burke}.}
  \bibinfo{year}{2020}\natexlab{}.
\newblock \showarticletitle{What are meaningful social interactions in
  today’s media landscape? a cross-cultural survey}.
\newblock \bibinfo{journal}{\emph{Social Media+ Society}} \bibinfo{volume}{6},
  \bibinfo{number}{3} (\bibinfo{year}{2020}),
  \bibinfo{pages}{2056305120942888}.
\newblock


\bibitem[\protect\citeauthoryear{Luutonen et~al\mbox{.}}{Luutonen
  et~al\mbox{.}}{2008}]%
        {luutonen2008handmade}
\bibfield{author}{\bibinfo{person}{Marketta Luutonen} {et~al\mbox{.}}}
  \bibinfo{year}{2008}\natexlab{}.
\newblock \showarticletitle{Handmade memories}.
\newblock \bibinfo{journal}{\emph{Trames}} \bibinfo{volume}{12},
  \bibinfo{number}{3} (\bibinfo{year}{2008}), \bibinfo{pages}{331--341}.
\newblock


\bibitem[\protect\citeauthoryear{Markopoulos}{Markopoulos}{2009}]%
        {markopoulos2009design}
\bibfield{author}{\bibinfo{person}{Panos Markopoulos}.}
  \bibinfo{year}{2009}\natexlab{}.
\newblock \showarticletitle{A design framework for awareness systems}.
\newblock In \bibinfo{booktitle}{\emph{Awareness systems}}.
  \bibinfo{publisher}{Springer}, \bibinfo{pages}{49--72}.
\newblock


\bibitem[\protect\citeauthoryear{Minowa and Gould}{Minowa and Gould}{1999}]%
        {minowa1999love}
\bibfield{author}{\bibinfo{person}{Yuko Minowa} {and}
  \bibinfo{person}{Stephen~J Gould}.} \bibinfo{year}{1999}\natexlab{}.
\newblock \showarticletitle{Love my gift, love me or is it love me, love my
  gift: A study of the cultural construction of romantic gift giving among
  Japanese couples}.
\newblock \bibinfo{journal}{\emph{ACR North American Advances}}
  (\bibinfo{year}{1999}).
\newblock


\bibitem[\protect\citeauthoryear{Monastero and McGookin}{Monastero and
  McGookin}{2018}]%
        {monastero2018traces}
\bibfield{author}{\bibinfo{person}{Beatrice Monastero} {and}
  \bibinfo{person}{David~K McGookin}.} \bibinfo{year}{2018}\natexlab{}.
\newblock \showarticletitle{Traces: Studying a public reactive floor-projection
  of walking trajectories to support social awareness}. In
  \bibinfo{booktitle}{\emph{Proceedings of the 2018 CHI Conference on Human
  Factors in Computing Systems}}. \bibinfo{pages}{1--13}.
\newblock


\bibitem[\protect\citeauthoryear{Monroy-Hern{\'a}ndez, Hill, Gonzalez-Rivero,
  and Boyd}{Monroy-Hern{\'a}ndez et~al\mbox{.}}{2011}]%
        {monroy2011computers}
\bibfield{author}{\bibinfo{person}{Andr{\'e}s Monroy-Hern{\'a}ndez},
  \bibinfo{person}{Benjamin~Mako Hill}, \bibinfo{person}{Jazmin
  Gonzalez-Rivero}, {and} \bibinfo{person}{Danah Boyd}.}
  \bibinfo{year}{2011}\natexlab{}.
\newblock \showarticletitle{Computers can't give credit: How automatic
  attribution falls short in an online remixing community}. In
  \bibinfo{booktitle}{\emph{Proceedings of the SIGCHI Conference on Human
  Factors in Computing Systems}}. \bibinfo{pages}{3421--3430}.
\newblock


\bibitem[\protect\citeauthoryear{Riche, Henry~Riche, Isenberg, and
  Bezerianos}{Riche et~al\mbox{.}}{2010}]%
        {riche2010hard}
\bibfield{author}{\bibinfo{person}{Yann Riche}, \bibinfo{person}{Nathalie
  Henry~Riche}, \bibinfo{person}{Petra Isenberg}, {and}
  \bibinfo{person}{Anastasia Bezerianos}.} \bibinfo{year}{2010}\natexlab{}.
\newblock \showarticletitle{Hard-to-use interfaces considered beneficial (some
  of the time)}.
\newblock In \bibinfo{booktitle}{\emph{CHI'10 Extended Abstracts on Human
  Factors in Computing Systems}}. \bibinfo{pages}{2705--2714}.
\newblock


\bibitem[\protect\citeauthoryear{Rosner and Ryokai}{Rosner and Ryokai}{2010}]%
        {rosner2010spyn}
\bibfield{author}{\bibinfo{person}{Daniela~K Rosner} {and}
  \bibinfo{person}{Kimiko Ryokai}.} \bibinfo{year}{2010}\natexlab{}.
\newblock \showarticletitle{Spyn: augmenting the creative and communicative
  potential of craft}. In \bibinfo{booktitle}{\emph{Proceedings of the SIGCHI
  conference on human factors in computing systems}}.
  \bibinfo{pages}{2407--2416}.
\newblock


\bibitem[\protect\citeauthoryear{Saquib, Kazi, Wei, and Li}{Saquib
  et~al\mbox{.}}{2019}]%
        {saquib2019interactive}
\bibfield{author}{\bibinfo{person}{Nazmus Saquib},
  \bibinfo{person}{Rubaiat~Habib Kazi}, \bibinfo{person}{Li-Yi Wei}, {and}
  \bibinfo{person}{Wilmot Li}.} \bibinfo{year}{2019}\natexlab{}.
\newblock \showarticletitle{Interactive body-driven graphics for augmented
  video performance}. In \bibinfo{booktitle}{\emph{Proceedings of the 2019 CHI
  Conference on Human Factors in Computing Systems}}. \bibinfo{pages}{1--12}.
\newblock


\bibitem[\protect\citeauthoryear{Scholz and Smith}{Scholz and Smith}{2016}]%
        {scholz2016immersive}
\bibfield{author}{\bibinfo{person}{Joachim Scholz} {and}
  \bibinfo{person}{Andrew~N. Smith}.} \bibinfo{year}{2016}\natexlab{}.
\newblock \showarticletitle{Augmented reality: Designing immersive experiences
  that maximize consumer engagement}.
\newblock \bibinfo{journal}{\emph{Business Horizons}} \bibinfo{volume}{59},
  \bibinfo{number}{2} (\bibinfo{year}{2016}), \bibinfo{pages}{149--161}.
\newblock
\showISSN{0007-6813}
\urldef\tempurl%
\url{https://doi.org/10.1016/j.bushor.2015.10.003}
\showDOI{\tempurl}


\bibitem[\protect\citeauthoryear{Scissors, Burke, and Wengrovitz}{Scissors
  et~al\mbox{.}}{2016}]%
        {scissors2016s}
\bibfield{author}{\bibinfo{person}{Lauren Scissors}, \bibinfo{person}{Moira
  Burke}, {and} \bibinfo{person}{Steven Wengrovitz}.}
  \bibinfo{year}{2016}\natexlab{}.
\newblock \showarticletitle{What's in a Like? Attitudes and behaviors around
  receiving Likes on Facebook}. In \bibinfo{booktitle}{\emph{Proceedings of the
  19th acm conference on computer-supported cooperative work \& social
  computing}}. \bibinfo{pages}{1501--1510}.
\newblock


\bibitem[\protect\citeauthoryear{Song, Vivrekar, Yeom, Paulos, and Salehi}{Song
  et~al\mbox{.}}{2021}]%
        {song2021crank}
\bibfield{author}{\bibinfo{person}{Katherine~W Song}, \bibinfo{person}{Janaki
  Vivrekar}, \bibinfo{person}{Lynn Yeom}, \bibinfo{person}{Eric Paulos}, {and}
  \bibinfo{person}{Niloufar Salehi}.} \bibinfo{year}{2021}\natexlab{}.
\newblock \showarticletitle{Crank That Feed: A Physical Intervention for Active
  Twitter Users}. In \bibinfo{booktitle}{\emph{Extended Abstracts of the 2021
  CHI Conference on Human Factors in Computing Systems}}.
  \bibinfo{pages}{1--6}.
\newblock


\bibitem[\protect\citeauthoryear{Spence}{Spence}{2019}]%
        {spence2019inalienability}
\bibfield{author}{\bibinfo{person}{Jocelyn Spence}.}
  \bibinfo{year}{2019}\natexlab{}.
\newblock \showarticletitle{Inalienability: Understanding digital gifts}. In
  \bibinfo{booktitle}{\emph{Proceedings of the 2019 CHI Conference on Human
  Factors in Computing Systems}}. \bibinfo{pages}{1--12}.
\newblock


\bibitem[\protect\citeauthoryear{Spottswood and Wohn}{Spottswood and
  Wohn}{2019}]%
        {spottswood2019beyond}
\bibfield{author}{\bibinfo{person}{Erin Spottswood} {and}
  \bibinfo{person}{Donghee~Yvette Wohn}.} \bibinfo{year}{2019}\natexlab{}.
\newblock \showarticletitle{Beyond the “like”: How people respond to
  negative posts on Facebook}.
\newblock \bibinfo{journal}{\emph{Journal of broadcasting \& electronic media}}
  \bibinfo{volume}{63}, \bibinfo{number}{2} (\bibinfo{year}{2019}),
  \bibinfo{pages}{250--267}.
\newblock


\bibitem[\protect\citeauthoryear{Stafford and Canary}{Stafford and
  Canary}{1991}]%
        {stafford1991maintenance}
\bibfield{author}{\bibinfo{person}{Laura Stafford} {and}
  \bibinfo{person}{Daniel~J Canary}.} \bibinfo{year}{1991}\natexlab{}.
\newblock \showarticletitle{Maintenance strategies and romantic relationship
  type, gender and relational characteristics}.
\newblock \bibinfo{journal}{\emph{Journal of Social and Personal
  relationships}} \bibinfo{volume}{8}, \bibinfo{number}{2}
  (\bibinfo{year}{1991}), \bibinfo{pages}{217--242}.
\newblock


\bibitem[\protect\citeauthoryear{Strauss and Corbin}{Strauss and
  Corbin}{1998}]%
        {strauss1998basics}
\bibfield{author}{\bibinfo{person}{A Strauss} {and} \bibinfo{person}{J
  Corbin}.} \bibinfo{year}{1998}\natexlab{}.
\newblock \showarticletitle{Basics of qualitative research techniques}.
\newblock  (\bibinfo{year}{1998}).
\newblock
\urldef\tempurl%
\url{https://doi.org/10.4135/9781452230153}
\showDOI{\tempurl}


\bibitem[\protect\citeauthoryear{Sugiyama}{Sugiyama}{2015}]%
        {sugiyama2015kawaii}
\bibfield{author}{\bibinfo{person}{Satomi Sugiyama}.}
  \bibinfo{year}{2015}\natexlab{}.
\newblock \showarticletitle{Kawaii meiru and Maroyaka neko: Mobile emoji for
  relationship maintenance and aesthetic expressions among Japanese teens}.
\newblock \bibinfo{journal}{\emph{First Monday}} (\bibinfo{year}{2015}).
\newblock


\bibitem[\protect\citeauthoryear{Vega}{Vega}{2017}]%
        {vega_2017}
\bibfield{author}{\bibinfo{person}{Nick Vega}.}
  \bibinfo{year}{2017}\natexlab{}.
\newblock \bibinfo{title}{I just lost a 159-day Snapchat streak and I couldn't
  be happier}.
\newblock
\newblock
\urldef\tempurl%
\url{https://www.businessinsider.com/snapchat-streak-lost-couldnt-be-happier-2017-8}
\showURL{%
\tempurl}


\bibitem[\protect\citeauthoryear{West, Quigley, and Kay}{West
  et~al\mbox{.}}{2007}]%
        {memento2007}
\bibfield{author}{\bibinfo{person}{David West}, \bibinfo{person}{Aaron
  Quigley}, {and} \bibinfo{person}{Judy Kay}.} \bibinfo{year}{2007}\natexlab{}.
\newblock \showarticletitle{MEMENTO: A Digital-Physical Scrapbook for Memory
  Sharing}.
\newblock \bibinfo{journal}{\emph{Personal Ubiquitous Comput.}}
  \bibinfo{volume}{11}, \bibinfo{number}{4} (\bibinfo{date}{apr}
  \bibinfo{year}{2007}), \bibinfo{pages}{313–328}.
\newblock
\showISSN{1617-4909}
\urldef\tempurl%
\url{https://doi.org/10.1007/s00779-006-0090-7}
\showDOI{\tempurl}


\bibitem[\protect\citeauthoryear{Wilson and Shpall}{Wilson and Shpall}{2016}]%
        {sep-action}
\bibfield{author}{\bibinfo{person}{George Wilson} {and} \bibinfo{person}{Samuel
  Shpall}.} \bibinfo{year}{2016}\natexlab{}.
\newblock \showarticletitle{{Action}}.
\newblock In \bibinfo{booktitle}{\emph{The {Stanford} Encyclopedia of
  Philosophy} (\bibinfo{edition}{{W}inter 2016} ed.)},
  \bibfield{editor}{\bibinfo{person}{Edward~N. Zalta}} (Ed.).
  \bibinfo{publisher}{Metaphysics Research Lab, Stanford University}.
\newblock


\bibitem[\protect\citeauthoryear{Wohn, Carr, and Hayes}{Wohn
  et~al\mbox{.}}{2016}]%
        {wohn2016affective}
\bibfield{author}{\bibinfo{person}{Donghee~Yvette Wohn},
  \bibinfo{person}{Caleb~T Carr}, {and} \bibinfo{person}{Rebecca~A Hayes}.}
  \bibinfo{year}{2016}\natexlab{}.
\newblock \showarticletitle{How affective is a “Like”?: The effect of
  paralinguistic digital affordances on perceived social support}.
\newblock \bibinfo{journal}{\emph{Cyberpsychology, Behavior, and Social
  Networking}} \bibinfo{volume}{19}, \bibinfo{number}{9}
  (\bibinfo{year}{2016}), \bibinfo{pages}{562--566}.
\newblock


\bibitem[\protect\citeauthoryear{Wolfinbarger}{Wolfinbarger}{1990}]%
        {wolfinbarger1990motivations}
\bibfield{author}{\bibinfo{person}{Mary~Finley Wolfinbarger}.}
  \bibinfo{year}{1990}\natexlab{}.
\newblock \showarticletitle{Motivations and symbolism in gift-giving behavior}.
\newblock \bibinfo{journal}{\emph{ACR North American Advances}}
  (\bibinfo{year}{1990}).
\newblock


\bibitem[\protect\citeauthoryear{Z{\"u}nd, Ryffel, Magnenat, Marra, Nitti,
  Kapadia, Noris, Mitchell, Gross, and Sumner}{Z{\"u}nd et~al\mbox{.}}{2015}]%
        {zund2015augmented}
\bibfield{author}{\bibinfo{person}{Fabio Z{\"u}nd}, \bibinfo{person}{Mattia
  Ryffel}, \bibinfo{person}{St{\'e}phane Magnenat}, \bibinfo{person}{Alessia
  Marra}, \bibinfo{person}{Maurizio Nitti}, \bibinfo{person}{Mubbasir Kapadia},
  \bibinfo{person}{Gioacchino Noris}, \bibinfo{person}{Kenny Mitchell},
  \bibinfo{person}{Markus Gross}, {and} \bibinfo{person}{Robert~W Sumner}.}
  \bibinfo{year}{2015}\natexlab{}.
\newblock \showarticletitle{Augmented creativity: Bridging the real and virtual
  worlds to enhance creative play}.
\newblock In \bibinfo{booktitle}{\emph{SIGGRAPH Asia 2015 Mobile Graphics and
  Interactive Applications}}. \bibinfo{pages}{1--7}.
\newblock


\end{thebibliography}

\end{document}